\documentclass[aps,pra,twocolumn,superscriptaddress,10pt,article,nofootinbib,showpacs,longbibliography]{revtex4-1}

\usepackage{amsfonts}
\usepackage{amsmath}
\usepackage{amsthm}
\usepackage{braket}
\usepackage{graphicx}
\usepackage[colorlinks=true,citecolor=blue,linkcolor=red,urlcolor=blue]{hyperref}
\usepackage{color}
\usepackage{amssymb}
\usepackage{dsfont}
\usepackage{verbatim}
\usepackage{amsmath,amscd}
\usepackage{multirow}

\usepackage[normalem]{ulem}
\usepackage{bm}
\usepackage[font=footnotesize]{caption}
\usepackage{subcaption}

\def \k{{\mathbf k}}
\def \q{{\mathbf q}}
\def \g{{\mathbf g}}
\def \X{{\mathbf X}}
\def \R{{\mathbf R}}
\def \r{{\mathbf r}}
\def \bdelta{{\boldsymbol\delta}}
\def \e{{\mathbf e}}
\def \u{{\mathbf u}}

\def \B{{\mathbf B}}
\def \K{{\mathbf K}}

\def \G{{\mathbf G}}

\def \n{{\mathbf n}}
\def \r{{\mathbf r}}
\def \bOmega{{\mathbf \Omega}}
\def \S{{\mathbf S}}

\def \s{{\mathbf s}}

\def \beq{\begin{eqnarray}}
\def \eeq{\end{eqnarray}}
\def \bro{\bar{\rho}_s}
\def \rp{{\bm r^{\prime}}}
\def \hx{\hat{x}}
\def \hz{\hat{z}}
\newcommand{\nn}{\nonumber \\}

\DeclareMathOperator{\Tr}{Tr}

\begin{document}

\title{Symmetry breaking and skyrmionic transport in twisted bilayer graphene}

\author{Shubhayu Chatterjee}
\author{Nick Bultinck}
\author{Michael P. Zaletel}
\affiliation{Department of Physics, University of California, Berkeley, California 94720, USA}

\begin{abstract}
Motivated by recent low-temperature magnetoresistance measurements in twisted bilayer graphene aligned with hexagonal boron-nitride substrate, we perform a systematic study of possible symmetry breaking orders in this device at a filling of two electrons per moir\'e unit cell. We find that the surprising non-monotonic dependence of the resistance on an out-of-plane magnetic field is difficult to reconcile with particle-hole charge carriers from the low-energy bands in symmetry broken phases. We invoke the non-zero Chern numbers of the twisted bilayer graphene flat bands to argue that skyrmion textures provide an alternative for the dominant charge carriers. Via an effective field-theory for the spin degrees of freedom, we show that the effect of spin Zeeman splitting on the skyrmion excitations provides a possible explanation for the non-monotonic magnetoresistance. We suggest several experimental tests, including the functional dependence of the activation gap on the magnetic field, for our proposed correlated insulating states at different integer fillings. We also discuss possible exotic phases and quantum phase transitions that can arise via skyrmion-pairing on doping such an insulator. 
\end{abstract}

\maketitle

\section{Introduction}

A series of recent experimental breakthroughs has uncovered surprising and fascinating correlated electron phenomena in two-dimensional van der Waals moir\'e materials. Transport experiments on twisted bilayer graphene \cite{Cao,Yankowitz,Efetov}, ABC trilayer graphene on hexagonal boron-nitride (hBN) \cite{Chen,ChenSharpe}, and twisted double bilayer graphene \cite{LiuHao,CaoRodan,Shen} show evidence of insulating states around charge neutrality at electron fillings for which no single-particle band-gap is expected. To make the story even more interesting, superconducting domes flanking some of these insulating states were observed \cite{Cao2,Yankowitz,Efetov,LiuHao,Shen}. In Refs. \cite{Kerelsky,ChoiKemmer,JiangMao}, spatially resolved properties of the insulating states were studied using scanning tunneling microscopy (STM) experiments. Recently, transport experiments were also performed at larger temperatures, and revealed an interesting broad temperature range with a large and linearly increasing resistivity \cite{CaoChowdhury,Polshyn}.

The origin of the insulating and superconducting states can be traced back to the presence of bands with vanishing bandwidth in the mini- or moir\'e Brillouin zone. In twisted bilayer graphene (tBLG), such flat mini-bands were predicted to occur at special `magic' twist angles between the top and bottom graphene layer \cite{Bistritzer}; an exact flat band criterion was later obtained in Ref.~\cite{Tarnopolsky} for a chiral approximation of the tBLG continuum model \cite{Bistritzer,Lopes,CastroNeto}. 
In ABC trilayer graphene on hBN and twisted double bilayer graphene, similar flat mini-bands around charge neutrality can be obtained by applying a suitable displacement field \cite{Chen,YaHuiZhang1,LeeKhalaf}. Interestingly, the flat bands often also have non-trivial topological properties. For instance in tBLG, the flat bands have non-trivial fragile topology protected by the space group symmetries \cite{Po,Zou,SongWang,Hejazi}. In devices which have isolated flat bands, one generally finds broad parameter regimes where these bands have non-zero Chern number \cite{YaHuiZhang1,LiuDai,Xie,AHpaper,YaHuiZhang3,LeeKhalaf}.

In this work, we focus on flat bands which have a gap at the charge neutrality point (CNP). This is motivated by the experiments of Refs. \cite{Goldhaber,Serlin}, where the Dirac cones in the tBLG flat bands are gapped by the $C_{2v}$ symmetry breaking AB-sublattice splitting induced by the hBN substrate. Although we focus on the case where the bandgap at charge neutrality has a trivial single-particle origin, most of our results can also be applied to mean-field band structures where the gap at the CNP results from spontaneous symmetry breaking induced by electron interactions. In tBLG, $C_{2v}T$ symmetry (with $T$ being time-reversal) needs to be spontaneously broken in order to generate a mean-field gap at charge neutrality. Self-consistent Hartree-Fock studies have found that this indeed happens for certain interaction strengths and twist angles \cite{Xie,Efetov,LiuKhalaf}. It was found that the $C_{2v}T$ symmetry breaking self-consistent Hartree-Fock solutions are very susceptible to $C_{3v}$ breaking strain \cite{LiuKhalaf}, an observation which agrees with the STM and transport experiments \cite{ZhangPo}.

Our main focus is tBLG with a single-particle gap at charge neutrality at electron filling $\nu=2$, i.e. at a doping of two electrons per moir\'e unit cell with respect to charge neutrality. Based on a phenomenological mean-field analysis, we argue that the magnetoresistance measurements of Ref. \cite{Goldhaber} impose very non-trivial constraints on the state that is realized at $\nu=2$. We analyse the different possible symmetry breaking orders and find that (almost) all of them are hard to reconcile with the transport measurements of Ref. \cite{Goldhaber}, given that we assume the charge carriers to be conventional particle-hole excitations. However, because of the non-trivial topology of the flat bands, skyrmions textures in a spin-polarized flat band carry electric charge \cite{Sondhi}. We study the potential role of skyrmions as the dominant charge carriers and find that they provide a natural explanation of
the experimental data of Refs. \cite{Goldhaber,Serlin}. We therefore posit that skyrmions contribute to transport in tBLG, and we provide experimentally falsifiable predictions for the activation gap as a function of out-of-plane magnetic field for insulators at $\nu = 2, 3$ to test our assertion. Towards the end of the manuscript, we speculate on skyrmion-pairing and possible connections to superconductivity.

\section{Magic-angle twisted bilayer graphene aligned with hBN}

We consider tBLG at the first magic angle $\theta\approx 1.05^\circ$ \cite{Bistritzer}, encapsulated on both sides by a hBN substrate. If hBN is sufficiently aligned with graphene, it induces a non-negligible sublattice splitting $\Delta \, \sigma^z$, which results in a $C_{2}T$ breaking mass term at the Dirac points \cite{Jung,SanJose,Hunt13,Amet13}. Further, because of the mismatch in lattice constant between graphene and hBN, a second moir\'e pattern arises \cite{Jung2}. As the rotation angle between graphene and hBN decreases, both the induced Dirac mass term and the strength of the second moir\'e pattern increase. There is a regime where the hBN induced moir\'e pattern can be neglected, while there is nonetheless a sizable Dirac mass. Here, we consider the situation where the top graphene layer and hBN substrate are in this regime, while the bottom graphene layer is sufficiently unaligned with hBN and is therefore not affected by the substrate. We will often use a hBN induced sublattice splitting of $15$ meV, which is expected to be a good estimate based on the findings of Ref. \cite{Kim2018}. We refer to Appendix \ref{app:MH} for a detailed discussion of the moir\'e Hamiltonian used in this work.

\begin{figure}
\centering
\begin{subfigure}[t]{0.23\textwidth}
\includegraphics[width=\textwidth]{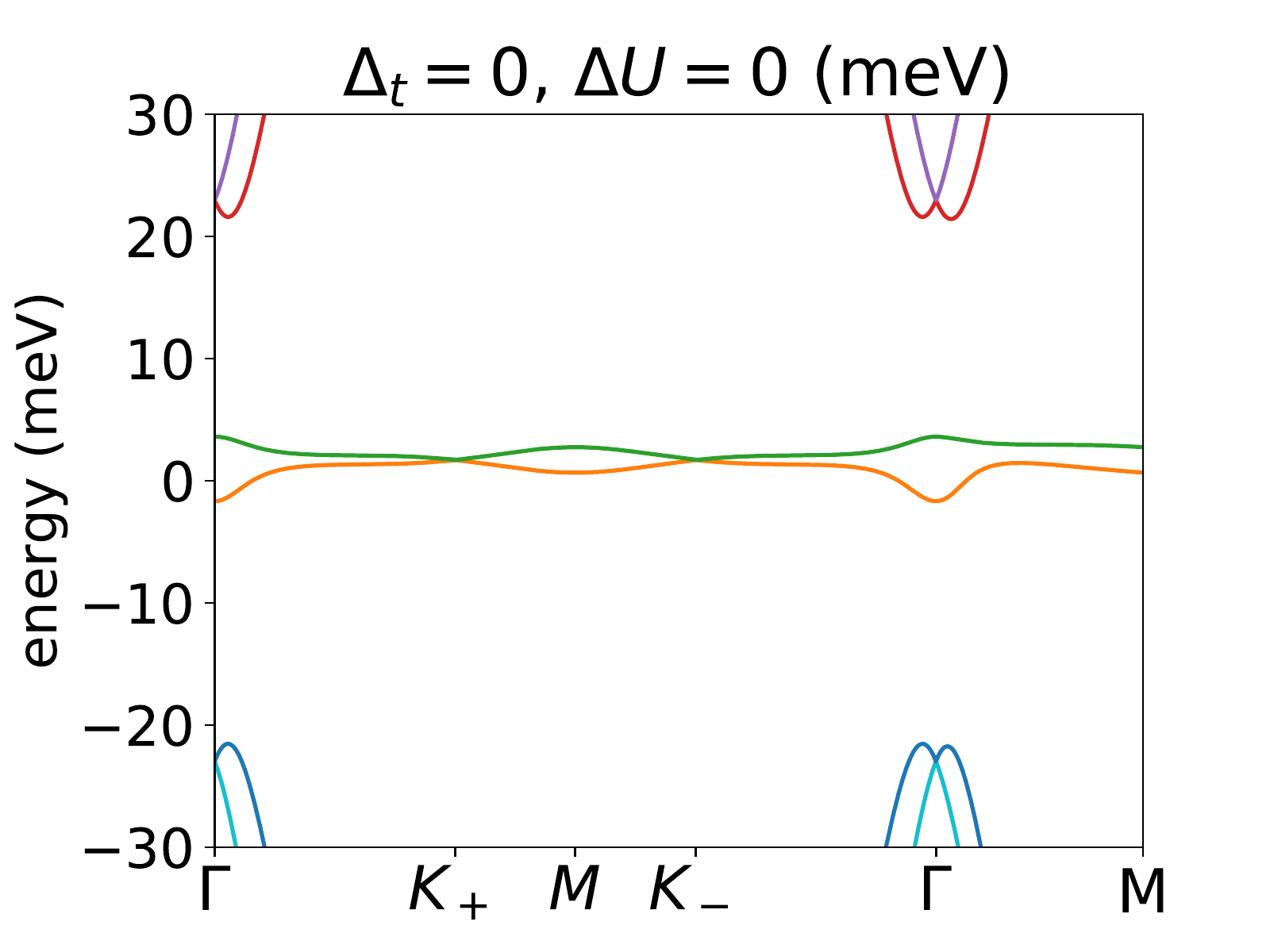}\caption{}
\end{subfigure}
\begin{subfigure}[t]{0.23\textwidth}
\includegraphics[width=\textwidth]{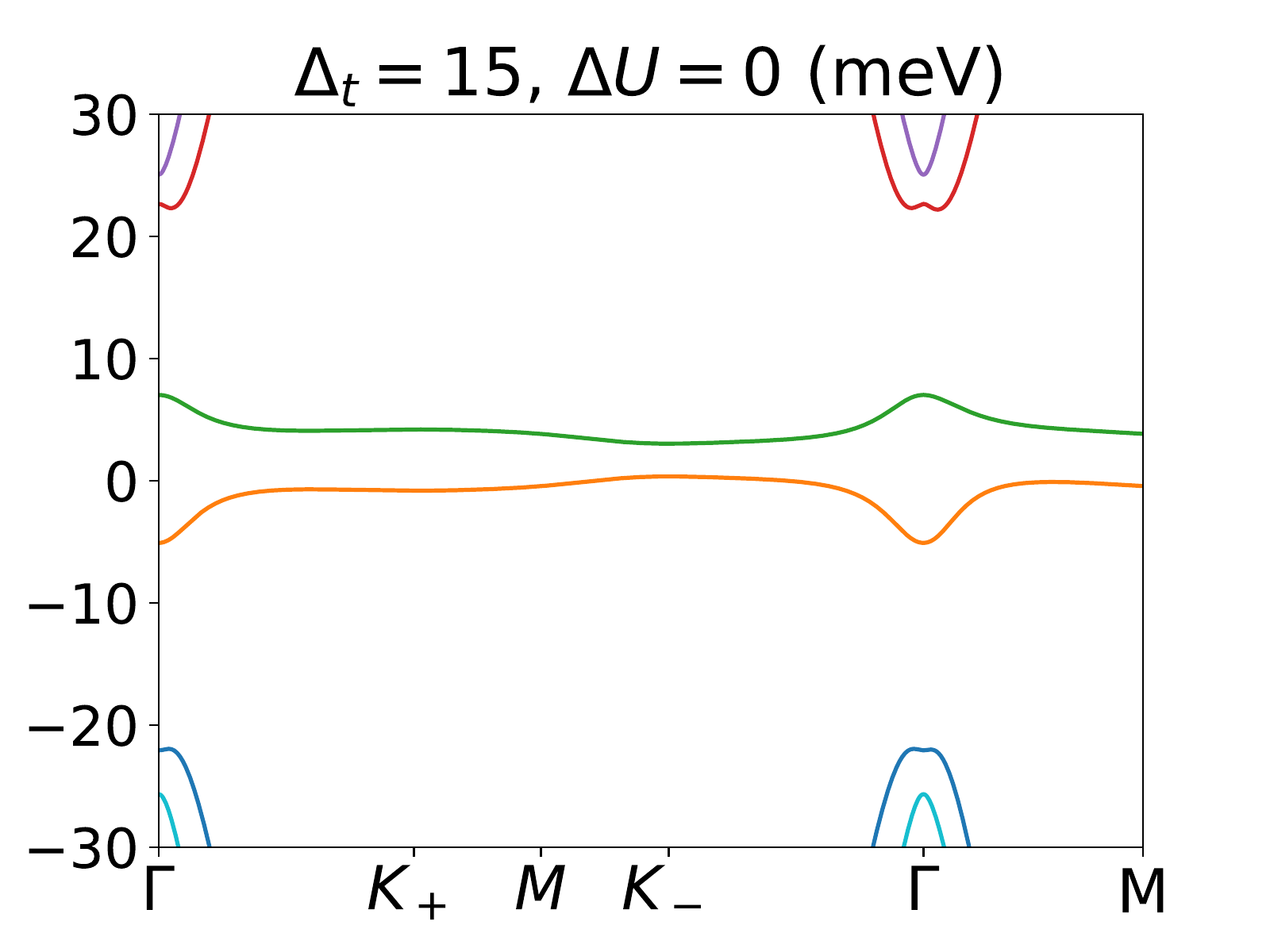}\caption{}
\end{subfigure}
\begin{subfigure}[t]{0.48\textwidth}
\includegraphics[width=\textwidth]{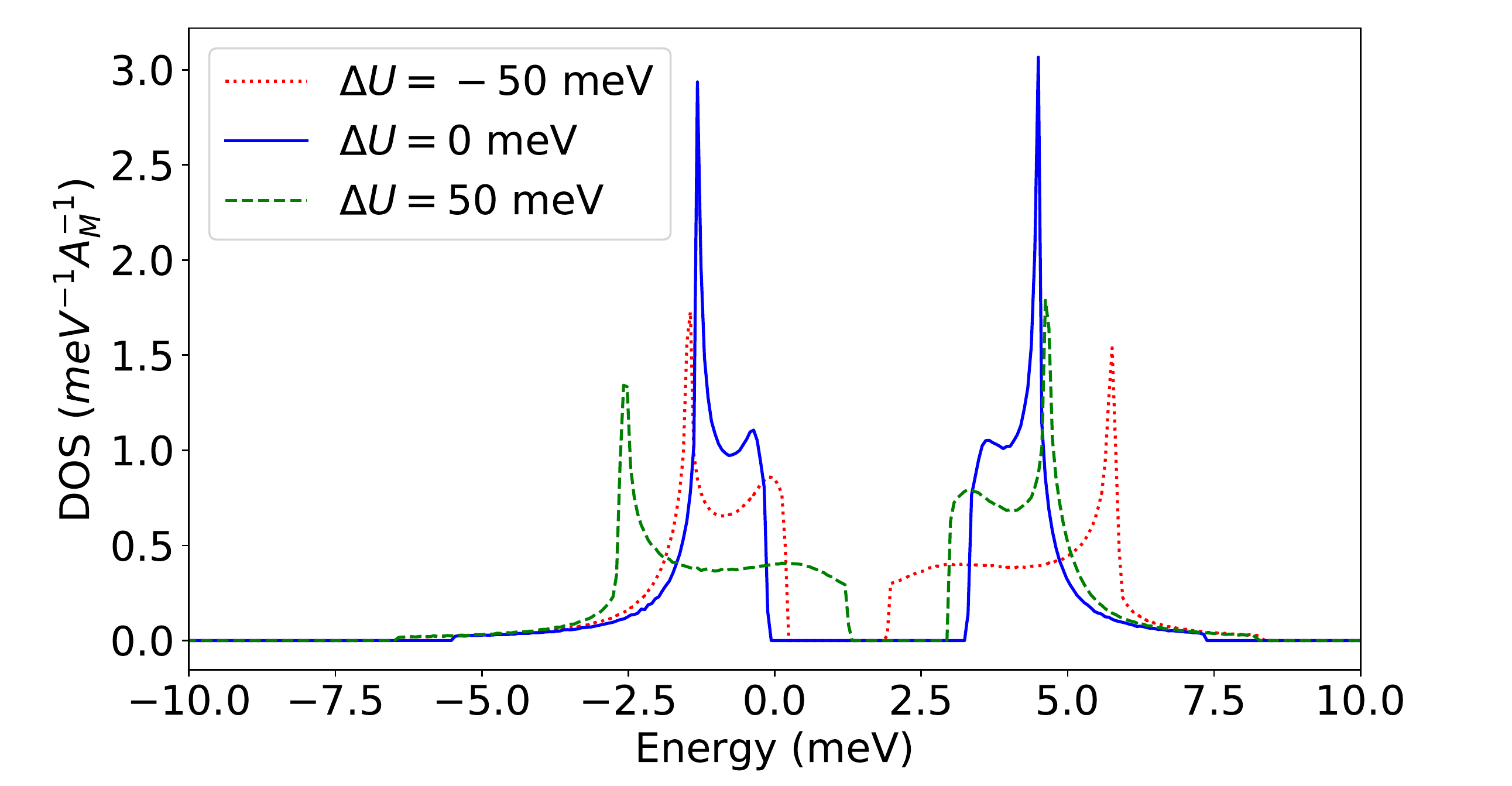}\caption{}
\end{subfigure}
\caption{(\textbf{a}) Band spectrum around charge neutrality of the single-valley tBLG moir\'e Hamiltonian at the first magic angle $\theta\approx 1.05^\circ$. At the K points in the mini-Brillouin zones, Dirac cones protected by $C_{2v}T$ are present. (\textbf{b}) With a sublattice splitting $\Delta_t$ of $15$ meV on the top graphene layer, induced by alignment with the hBN substrate, the Dirac cones acquire a mass. The resulting isolated valence and conduction bands carry non-zero Chern number $|C|=1$. (\textbf{c}) The effect on the flat band density-of-states (DOS) of a potential energy difference $\Delta U$ between top and bottom graphene layers as a result of non-zero displacement field, for $\Delta_t = 15$ meV. $A_M$ is the area of the moir\'e unit cell. The valence (conduction) band DOS is strongly affected by positive (negative) $\Delta U$.} \label{fig:BS}
\end{figure}

In Refs. \cite{AHpaper,YaHuiZhang3}, it was found that a non-zero sublattice splitting on one side of magic-angle tBLG gaps out all Dirac cones of the moir\'e Hamiltonian. Because the two Dirac cones in a single-valley moir\'e Hamiltonian, shown in Fig. \ref{fig:BS}(a), originate from the two different graphene layers, this is a consequence of the inter-layer coupling. Ignoring spin, the single-valley moir\'e Hamiltonian with sublattice splitting on one layer has two isolated flat bands, as shown in Fig. \ref{fig:BS}(b). The Chern numbers of these bands were calculated in Refs. \cite{AHpaper,YaHuiZhang3}, and found to be $C=\pm 1$. Note that once we know the Chern number $C$ of one band, all the other Chern numbers are fixed. This is because the total Chern number in one valley always adds up to zero (as long as the sublattice splitting is not strong enough to mix the flat bands with the dispersive bands), and because the two valleys are interchanged by time-reversal symmetry, which changes the sign of the Chern number. With positive sublattice splitting $\Delta$ on one of the graphene layers, the band above charge neutrality in valley $+$, i.e. the valley at the $K$ points of the mono-layer graphene Brillouin zone, has $C= -1$.

In Refs.~\cite{Goldhaber,Serlin}, spontaneous time-reversal symmetry breaking at $\nu=3$ was observed in a magic-angle tBLG device where one of the graphene layers is nearly aligned with hBN. In particular, Ref. \cite{Goldhaber} reported a rotational mismatch between the top graphene layer and the hBN substrate of $\approx 0.83^\circ$. In both experiments, the spontaneous time-reversal breaking is accompanied by a non-zero anomalous Hall effect. On top of this, ref. \cite{Serlin} observed insulating behavior at $\nu=3$, and a corresponding quantized Hall conductance $\sigma_{xy}=\pm e^2/h$. Because of the non-zero Chern numbers of the flat bands with hBN alignment, these experimental observations at $\nu=3$ can be naturally explained if the Coulomb interactions cause the electrons to spontaneously polarize into one valley \cite{AHpaper,YaHuiZhang3}; complete spin polarization in addition to valley polarization can lead to an insulator with quantized $\sigma_{xy}$. In this work we focus on the experimental findings for the same devices at filling $\nu=2$. At this filling, no anomalous Hall effect was observed, but a clear resistance peak is nevertheless present \cite{Goldhaber,Serlin}. Although an activation gap is yet to be observed at $\nu=2$ in transport measurements, this resistance peak hints at the possibility of a true insulating state at zero temperature. Here we assume that such an insulating state is indeed realized at lower temperatures.

Before going into the interaction effects that stabilize the putative insulator at $\nu=2$, we first discuss one last single-particle effect. In Ref. \cite{Goldhaber}, it was observed that applying a displacement field along one direction destroys the resistance peak at $\nu=2$, while this peak is almost insensitive to a displacement field applied in the other direction. To understand this behavior, we studied the effect of a non-zero potential energy difference between top and bottom graphene layers on the flat bands. In Fig. \ref{fig:BS} (c), we show the density of states (DOS) of the flat bands with a sublattice splitting $\Delta_t=15$ meV, and a potential energy difference $\Delta U=U_t-U_b$ of $0$, $50$ and $-50$ meV. We see that for $\Delta U=50$ meV, there is only a small change in the conduction band DOS as compared to the case when $\Delta U=-50$ meV. Fig. \ref{fig:BS} clearly shows that for negative $\Delta U$, the conduction band DOS decreases more, and spreads over a larger energy window as  function of $|\Delta U|$. At the very least, this dependence of the DOS on displacement field, and in particular on the sign of $\Delta U$, is consistent with the scenario that the resistance peak at $\nu=2$ is attributed to a correlated insulator, because a lower DOS and a larger bandwidth reduce the effect of electron interactions.

\section{Possible symmetry breaking orders at $\nu=2$}

To address the nature of the correlated insulator observed at $\nu=2$, we follow the phenomenological approach of Ref. \cite{YaHuiZhang3} and identify the symmetry breaking orders that are compatible with the experimental observations (for simplicity, we neglect spatial symmetry breaking on the moir\'e scale). We note that recently a similar phenomenological approach was used to distinguish different pairing order parameters in tBLG and twisted double bilayer graphene \cite{WuDasSarma2,Scheurer}. The dominant terms in the Hamiltonian are U$(2)_+\times$U$(2)_-$ symmetric, where the $\pm$ subscript refers to the valley quantum number. The U$(2)_+\times$U$(2)_-$ symmetry consists of overall charge conservation, valley-charge conservation, and independent SU$(2)$ spin rotations in each valley. We write its corresponding Lie algebra as $\mathds{1},\tau^z,\s$ and $\tau^z\s$, where $\tau^i$ and $s^i$ are the Pauli matrices  acting respectively on the valley and spin indices. The total Hamiltonian also contains terms that break the SU$(2)_+\times$SU$(2)_-$ subgroup down to the physical SU$(2)$ spin rotation group, but they operate at much lower energy scales. We will ignore these terms for now, and discuss them in more detail in the next section. We can organize the fifteen order parameters $\tau^i s^j$ into three different multiplets under U$(2)_+\times$U$(2)_-$ \cite{YaHuiZhang3}: (1) $\tau^z$, (2) ($\tau^{x/y}, \tau^{x/y}\s$) and (3): ($\s, \tau^z\s$).

The order parameter $\tau^z$ corresponds to a spin singlet, valley-polarized insulator where all electrons occupy the same valley. This possibility can readily be excluded, since in this case the system would be an anomalous Hall insulator with $\sigma_{xy}=\pm 2e^2/h$. However, no sign of non-zero Hall conductivity at zero magnetic field was observed at $\nu=2$ \cite{Goldhaber}.

\begin{figure}
\centering 
\begin{subfigure}[t]{0.18\textwidth}
\includegraphics[width=\textwidth]{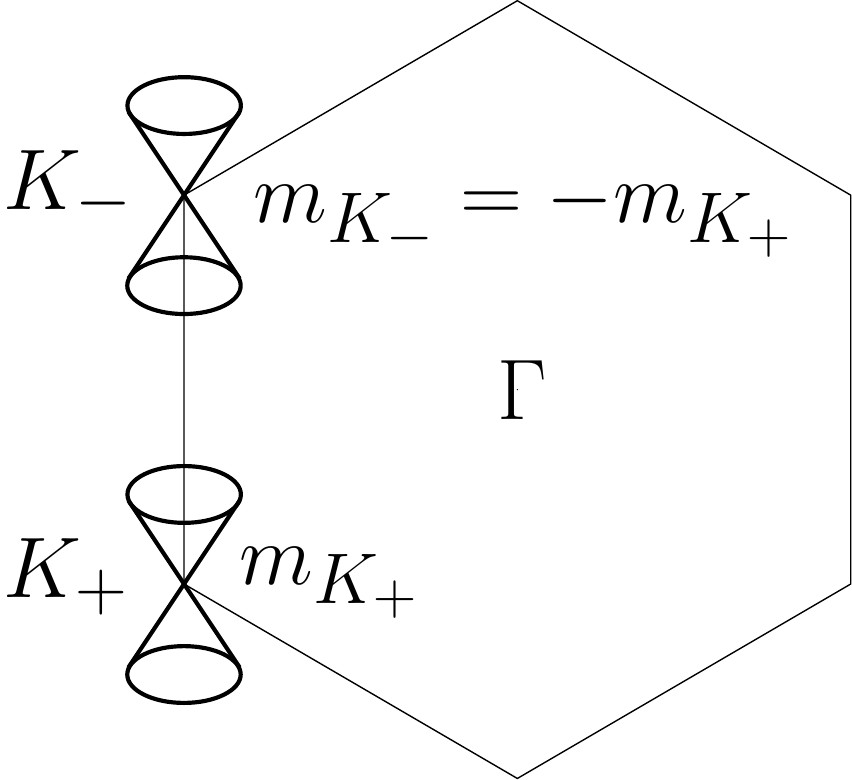}\caption{}
\end{subfigure}\hspace{0.5 cm}
\begin{subfigure}[t]{0.22\textwidth}
\includegraphics[width=\textwidth]{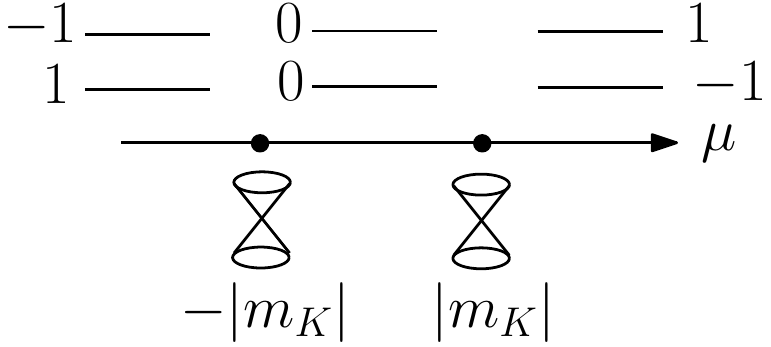}\caption{}
\end{subfigure}
\caption{(\textbf{a}) Mini-Brillouin zone with the Dirac cones at the $K_+$ and $K_-$ points coming from the IVC insulator order parameter $M^x(\k)\tau^x+M^y(\k)\tau^y$. Both Dirac cones have the same chirality. The mass terms at $K_+$ and $K_-$, which have opposite signs, come from the flat band dispersion: $m_{K_\pm}= \pm (\varepsilon_{+,\K_+}- \varepsilon_{+,\K_-})/2$. (\textbf{b}) Effect of a fictitious term $\mu\tau^z$ on the IVC insulator mean-field Hamiltonian. Tuning $\mu$ from minus infinity to plus infinity induces two Chern number changing transitions, where the Chern number of the valence (conduction) band changes from $1$ ($-1$) to $0$, and from $0$ to $-1$ ($1$) (for positive $\Delta_t$). In the figure, above the $\mu$ axis we schematically show the valence and conduction bands with their respective Chern number.}\label{fig:cones}
\end{figure}

The second possibility is that the ground state corresponds to an intervalley coherent (IVC) state, with order parameter multiplet ($\tau^{x/y}, \tau^{x/y}\s$). Let us pick the $\tau^x, \tau^y$ order parameters, and write the mean field Hamiltonian for the four bands above charge neutrality (including spin) as $H_{MF} = \sum_{\k} c^{\dagger}_{\k, \tau, s} [h_{\k}]_{\tau,s; \tau^\prime, s^\prime} c_{\k, \tau^\prime, s^\prime}$, where $\k$ lies in the mini-Brillouin zone (MBZ). For the IVC state, restricting to an out-of-plane magnetic field ($B_\parallel = 0$),

\beq \label{ham1}
h_\k &=& \frac{(\varepsilon_{+,\k}-\varepsilon_{-,\k})}{2} \tau^z \otimes s^0 +M^x(\k)\tau^x \otimes s^0 +
M^y(\k)\tau^y \otimes s^0 \nonumber\\ && -\frac{\mu_B g_v(\k)B_\perp}{2}\tau^z \otimes s^0 - \frac{\mu_B g_s B_\perp}{2} \tau^0 \otimes s^z\, ,
\eeq
where $\varepsilon_{\tau,\k}$ is the band energy in valley $\tau$. Note that we have dropped an unimportant term proportional to the identity. The first term on the second line in Eq. \eqref{ham1} is the valley Zeeman term, with $\mu_B$ the Bohr magneton, which describes the coupling between an out-of-plane magnetic field $B_\perp$ and the orbital magnetic moment of the electrons \cite{Thonhauser,XiaoNiu,Xiao}. The last term is the conventional spin-Zeeman term. Time reversal acts on the Hamiltonian in Eq. \eqref{ham1} as $\tau^xK$, where $K$ means complex conjugation. Let us first analyze this mean field Hamiltonian for $B_\perp=0$. Because the flat bands above charge neutrality have Chern number $C=\pm 1$, we know that $M(\k)=M^x(\k)+iM^y(\k)$ has at least two nodes in the mini-Brillouin zone with the same phase winding \cite{Murakami,AHpaper} (see also \cite{Dukan}). Assuming the minimal scenario with only two nodes is realized, $C_{3v}$ and time-reversal symmetry dictate that these nodes are located either at the $K_+$ and $K_-$ points of the mini-Brillouin zone, or both at the $\Gamma$ point. Since the IVC mass $M$ set by the Coulomb scale ($\approx 20$ meV) is expected to be much larger than the non-interacting bandwidth ($\approx 3$ meV), the minimum band gap corresponds to the nodes of $M(\k)$ in the MBZ. Therefore, when the nodes are at the $K$ points, the band gap of the mean field Hamiltonian is given by $|\varepsilon_{+,\K_+}-\varepsilon_{-,\K_+}| = |\varepsilon_{+,\K_+}-\varepsilon_{+,\K_-}|$, where we have used $\varepsilon_{-,\k}=\varepsilon_{+,-\k}$ as follows from time-reversal symmetry. We will refer to this possibility as the IVC insulator. If the nodes are both at $\Gamma$, then the mean field Hamiltonian is a semi-metal, which we will refer to as the IVC semi-metal. Let us first elaborate on the topological properties of the gapped bands of the IVC insulator. Because the nodes of $\Delta(\k)$ have the same winding, the resulting Dirac cones in the mean field Hamiltonian have the same chirality. The mass terms $m_{K_+}\tau^z$ and $m_{K_-}\tau^z$ at the $K_+$ and $K_-$ points coming from the flat-band dispersion have opposite sign, as can easily be seen from $m_{K_+} = (\varepsilon_{+,\K_+}-\varepsilon_{-,\K_+})/2 = (\varepsilon_{+,\K_+}-\varepsilon_{+,\K_-})/2$ and $m_{K_-} = (\varepsilon_{+,\K_-}-\varepsilon_{-,\K_-})/2 = (\varepsilon_{+,\K_-}-\varepsilon_{+,\K_+})/2$. So we conclude that the bands of the IVC insulator mean field Hamiltonian have zero Chern number. This can also be seen by adding a fictitious term $\mu\tau^z$ to the Hamiltonian in Eq.~\eqref{ham1}. Tuning $\mu$ from minus infinity to plus infinity induces two Chern number changing transitions, where at each transition the Chern number changes by one at a Dirac cone located at one of the nodes of $M(\k)$. This is shown schematically in Fig.~\ref{fig:cones} (b). 

\begin{figure}
    \centering
    \includegraphics[scale=0.55]{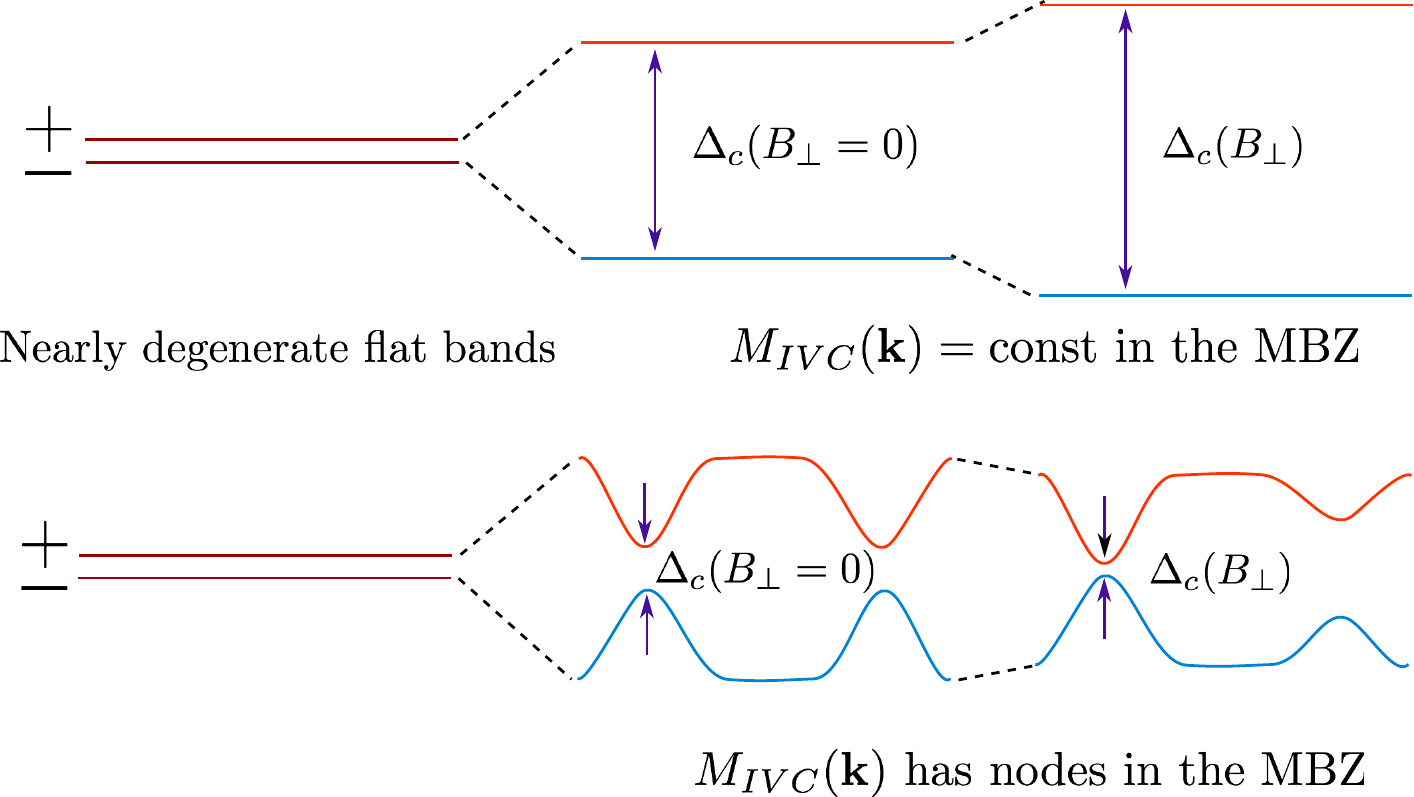}
    \caption{Schematic charge gap $\Delta_c$ as a function of $B_\perp$ for the IVC-I state, neglecting the small spin Zeeman effect. $\Delta_c$ would increase for uniform non-zero $M_{IVC}(\k)$. This is not allowed by the opposite Chern numbers of the two valleys and hence $\Delta_c$ decreases at one node of $M_{IVC}(\k)$.}
    \label{fig:bandSplittingsIVC}
\end{figure}

Now we investigate the consequences of turning on a non-zero out-of-plane magnetic field. We first consider the IVC insulator. For non-zero $B_\perp$, the valley-Zeeman term starts to compete with the mass terms $m_{K_+}\tau^z$ and $m_{K_-}\tau^z$. Since $m_{K_+} = -m_{K_-}$, the valley Zeeman effect must decrease the gap at either $K_+$ or $K_-$ (and increase the gap at the other point, see Fig.~\ref{fig:bandSplittingsIVC}), regardless of the sign of the perpendicular magnetic field. At the twist angle used in Ref. \cite{Goldhaber}, and with $\Delta_t=15$ meV, the magnitude of $g_v(\k)$ is approximately $15$ at the mini-Brillouin zone $K$ points \cite{YaHuiZhang3}. Because of this, we can safely ignore the spin-Zeeman term. From the mean-field Hamiltonian Eq. \eqref{ham1}, we see that the band gap of the IVC insulator is given by

\begin{equation}
\Delta_{IVC-I}(B_\perp)= 2|m_K|-\mu_B|g_v(\K)B_\perp|
\end{equation}
Irrespective of the sign of $B_\perp$, the band gap $\Delta_{IVC-I}$ closes when $\mu_B |g(\K)B_\perp|/2 = |m_{K}|$, where $|m_K| = |m_{K_+}| = |m_{K_-}|$.  Given that $|m_K|\approx 1.5$ meV, we find that the bandgap of the IVC insulator closes when $B_\perp \approx 3-4$ T. However, this behavior, schematically depicted in Fig.~\ref{fig:bandSplittingsIVC}, is difficult to reconcile with the experimental findings of Ref.~\cite{Goldhaber} as the magnetoresistance measurements show an increase in resistivity at $\nu=2$ as a function of out-of-plane magnetic field, with a resistance peak around $6$ T. 

\begin{figure}
\centering 
\begin{subfigure}[t]{0.12\textwidth}
\vspace{-2.0 cm}
\includegraphics[width=\textwidth]{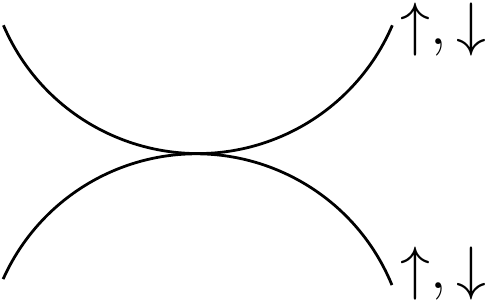}\vspace{0.63 cm}\caption{}
\end{subfigure}\hspace{0.4 cm}
\begin{subfigure}[t]{0.12\textwidth}
\includegraphics[width=\textwidth]{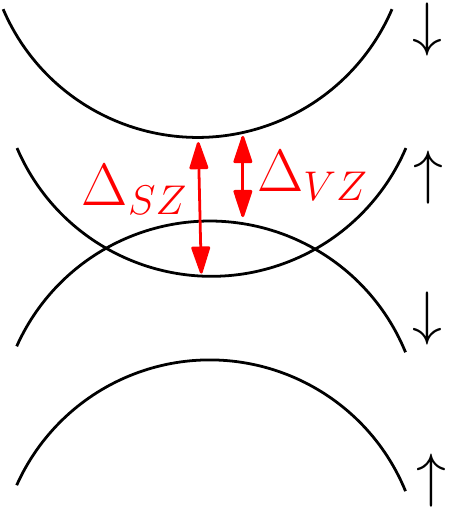}\caption{}
\end{subfigure}\hspace{0.4 cm}
\begin{subfigure}[t]{0.12\textwidth}
\includegraphics[width=\textwidth]{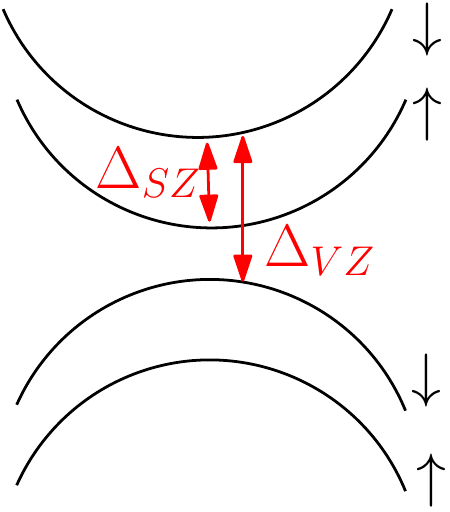}\caption{}
\end{subfigure}
\caption{Band spectrum around $\Gamma$ of the IVC semi-metal mean-field Hamiltonian, corresponding to Eq. \eqref{ham1} with both nodes of $M(\k)$ located at $\Gamma$. (\textbf{a}) Band spectrum at $B_\perp=0$. (\textbf{b}) With non-zero $B_{\perp}$, the band spectrum develops a Fermi surface if the spin-Zeeman splitting $\Delta_{SZ}=|g_s\mu_BB_\perp|$ is greater than the valley-Zeeman splitting $\Delta_{VZ}=|g_v(0)\mu_B B_\perp|$. (\textbf{c}) The opposite case compared to (b). Now the valley-Zeeman splitting is larger than the spin-Zeeman splitting, resulting in a gapped band spectrum.} \label{fig:quadr}
\end{figure}

For the IVC semi-metal, the valley-Zeeman term will generate a mass term at $\Gamma$. The spin-Zeeman term lifts the spin degeneracy, which makes the valence and conduction bands overlap around $\Gamma$. The net effect of the out-of-plane magnetic field depends on the sign of $g_s-g_v(0)$, where $g_v(0)$ is the orbital $g$-factor at $\Gamma$. As we show in Fig.~\ref{fig:quadr}, if the spin-Zeeman splitting $\Delta_{SZ}(B_\perp)=|g_s\mu_BB_\perp|$ is bigger than the valley-Zeeman splitting $\Delta_{VZ}(B_\perp)=|g_v(0)\mu_B B_\perp|$, a Fermi surface appears around $\Gamma$. If $\Delta_{VZ}(B_\perp)>\Delta_{SZ}(B_\perp)$, then the IVC semi-metal develops an energy gap at $\Gamma$. We find that $g_v(0)$ depends sensitively on twist angle, lattice relaxation and sublattice splitting. However, generically $g_v(0)>g_s$, such that an out-of-plane magnetic field creates a non-zero energy gap. The IVC semi-metal is thus consistent with the magnetoresistance measurements of Ref. \cite{Goldhaber}. However, we expect such a phase to be energetically unfavorable for two reasons. First, the Fermi surface is not entirely gapped out at $B_\perp = 0$, which means that the fermions gain less correlation energy compared to other order parameters that lead to a fully gapped spectrum. Second, a double vortex in $M(\k)$ costs twice the energy of two single vortices from a symmetry allowed term of the form $\int_\k |\nabla_\k M(\k)|^2$ in the effective action, as the latter endows a vortex with an energy cost proportional to the square of its winding number. Therefore, below we will focus on the possibility of an insulating state at $\nu=2$.

Let us also briefly comment on the possibility that $C_{3v}$ and/or time-reversal are spontaneously broken. In that case, the nodes of $M(\k)$ appear at generic positions in the mini-Brillouin zone, and will be gapped out by the mass terms $(\varepsilon_{+,\k}-\varepsilon_{-,\k})\tau^z/2$ at the locations of the nodes. For non-zero $B_\perp$, both the valley-Zeeman and the spin-Zeeman terms will compete with these mass terms, similar to the case when the nodes are at the $K$-points, as long as time-reversal is preserved and hence the gap decreases for either direction of $B_\perp$. However, if $M(\k)$ spontaneously breaks time-reversal and $C_{3v}$, it is possible for both mass terms to have the same sign at the location of the nodes. In this case, the band gap will decrease for one direction of $B_\perp$, but increase for the other direction. So this scenario could in principle explain the magnetoresistance measurements of Ref.~\cite{Goldhaber}, but it requires strong breaking of valley-U$(1)$, $C_{3v}$ and time-reversal. It can readily be identified in experiments by doing magnetoresistance measurements for both directions of $B_\perp$ and observing opposite behavior of $R_{xx}(B_\perp)$.

The third and final possibility is that the insulator has an order parameter in the multiplet ($\s, \tau^z\s$), in which case the electrons fill one spin-polarized band in each valley. Let us assume the order parameter is $s^z$, and write down a corresponding mean-field Hamiltonian:

\beq\label{ham2}
h_{\k} = && \frac{(\varepsilon_{+,\k}-\varepsilon_{-,\k})}{2}\tau^z \otimes s^0 + M_S \, \tau^0 \otimes s^z \nonumber \\ &&  -\frac{\mu_B g_v(\k)B_\perp}{2} \tau^z \otimes s^0 - \frac{\mu_Bg_sB_\perp}{2} \tau^0 \otimes s^z
\eeq
In this case, the valley-Zeeman term competes with the order parameter mass term $M_S \, s^z$, and the mean-field band gap is given by 
\begin{equation}
\Delta_{VI}\approx 2|M_S|-\mu_B|g_{v,max}B_\perp|\, ,
\end{equation}
where $g_{v,max}$ is the maximal value of $g_v(\k)$ in the mini-Brillouin zone. Note that we have assumed that $M_S$ is much larger than the bandwidth of the flat bands, although our conclusions below will also be valid without this assumption (as long as $M_S$ is bigger than the bandwidth). We have also ignored the spin-Zeeman term because the maximal orbital $g$-factor is much larger than the spin $g$-factor. The bandgap $\Delta_{VI}$ again decreases with an out-of-plane magnetic field. So at first sight, also this insulator seems incompatible \cite{YaHuiZhang3} with the experimental findings of Ref.~\cite{Goldhaber}. However, in contrast to the IVC insulator, now the bands of the mean-field Hamiltonian in Eq. \eqref{ham2} have Chern number $C=\pm 1$. It is well-known in the context of quantum Hall ferromagnetism \cite{Sondhi,Girvin,Eisenstein} that skyrmion textures in a spin-polarized Landau level carry electric charge \cite{Sondhi}. This is also true for Chern insulators, which means that there is another candidate for the lowest-energy charged excitations. If skyrmions are indeed the lowest-energy charge carriers, then the resistivity increase with out-of-plane magnetic field in the transport measurements of Ref.~\cite{Goldhaber} would result from the spin-Zeeman term, which increases the energy of a skyrmion. In the next sections, we examine this possibility in more detail. We note that skyrmions in general flat moir\'e bands with non-zero Chern number were also discussed in Ref.~\cite{YaHuiZhang1}. While Ref.~\cite{YaHuiZhang1} focuses on the possibility of skyrmionic superconductivity for bosonic skyrmions in $C=2$ Chern bands, in our work we mainly focus on fermionic skyrmions in $C=1$ bands and their implication on the gap. 

\section{SU$(2)_+\times$SU$(2)_-$ symmetry breaking effects}\label{symmbreaking}

In the previous section we have argued that if the resistance peak observed in Ref. \cite{Goldhaber} at $\nu=2$ can be attributed to an insulating state, then this insulator has a symmetry breaking order parameter in the multiplet $(\s,\tau^z\s)$, and skyrmions as lowest-energy charge carriers. Before discussing the skyrmion excitations in more detail, we first study the SU$(2)_+\times$SU$(2)_-$ symmetry breaking terms in the Hamiltonian, which distinguish between the $\s$ and $\tau\s$ order parameters. We want to know what order parameter gives the lowest energy, i.e. whether the SU$(2)_+\times$SU$(2)_-$ breaking terms favor spin alignment or anti-alignment between the different valleys. If the spins are aligned in the two valleys (order parameter $\s$), the insulator is a time-reversal symmetry breaking ferromagnet with a non-zero local spin moment. If the spins in the valleys are anti-aligned (order parameter $\tau^z\s$), the insulator is time-reversal symmetric which implies there is no local spin moment. Because the electron spin in this state is locked to the valley quantum number ($s^z= \frac{\tau}{2}$ or $s^z = -\frac{\tau}{2}$), we will refer to it as the `spin-valley locked state'. In a non-zero external magnetic field, the spins in the spin-valley locked insulator will cant in the direction of the magnetic field, similar to the canted anti-ferromagnet (CAF) \cite{SachdevSenthil,Zheng,Zheng2}. The canted spin-valley locked state which appears in this manuscript is similar to the CAF occuring in the $\nu=0$ graphene Landau levels \cite{Kharitonov1,Kharitonov,Pezzini,YoungSanchez}.

A first microscopic SU$(2)_+\times$SU$(2)_-$ breaking term comes from the Coulomb interaction, which takes the form
\begin{equation}\label{Coulomb}
H_C = \frac{1}{2A} \sum_{\q}   \sum_{l,l'} V_{ll'}(\q):\rho_{l}(\q)\rho_{l'}(-\q):\, ,
\end{equation}
where $l=t,b$ is a layer index and $A$ is the area of the mono-layer graphene unit cell. From now on we will always implicitly assume normal ordering. For the interaction potential we use a dual-gate screened Coulomb potential, which in momentum space takes the form
\begin{eqnarray}
\label{eq:ScrC}
V_{tt}(\q) = V_{bb}(\q) & = & \frac{e^2}{2\epsilon_r\epsilon_0 |\q|} \tanh(D|\q|) \\
V_{tb}(\q) = V_{bt}(\q) & = & \frac{e^2}{2\epsilon_r\epsilon_0 |\q|} \left(e^{-d|\q|}-\frac{2e^{-2D|\q|}}{1+e^{-2D|\q|}} \right)\label{eq:dD}
\end{eqnarray}
where $D$ is the distance from the tBLG to the metallic gates, which we take to be three moir\'e lattice constants. Eq.~\eqref{eq:dD} holds when the inter-layer distance $d$, of the order of one graphene lattice constant, is much smaller than the gate distance $D$. Based on the findings of Ref. \cite{Hunt}, we take the hBN dielectric constant to be $\epsilon_r = 6.6$. The layer resolved density operator $\rho_l(\q)$ is given by

\begin{equation}
\rho_l(\q) = \frac{1}{\sqrt{N}}\sideset{}{'}\sum_\k\sum_{\sigma,s} \psi^\dagger_{\k+\q,l,\sigma,s}\psi_{\k,l,\sigma,s}\, ,
\end{equation}
where $N$ is the number of graphene unit cells and $\sigma$ and $s$ are respectively sublattice and spin indices. We use primed momentum sums to denote sums that run over the mono-layer graphene Brillouin zone. A few remarks are in order before we proceed with our analysis. We have used the expression $V(\q)=\int \mathrm{d}^2\r\, V(\r)e^{i\q\cdot\r}$ for the interaction potential in Fourier space. This approximation is valid for $a|\q|\ll 1$, with $a$ the graphene lattice constant. However, the inter-valley scattering terms we are interested in involve large momentum transfers between electrons, and are therefore not in the regime where $a|\q|\ll 1$ holds. Although $V(\q)$ does not accurately describe lattice-scale interactions, we nevertheless still expect it to give a reliable estimate for the energy scale of the inter-valley scattering, and to provide the correct physical picture of the SU$(2)_+\times$SU$(2)_-$ symmetry breaking effects.

We now project the density operators in the flat bands above charge neutrality, which gives

\begin{equation}
\tilde{\rho}_{l,\g}(\q) = \frac{1}{\sqrt{N}}\sum_{\tau,\tau'}\sum_{\k}\lambda^{\tau,\tau'}_{l,\g}(\q,\k)c^\dagger_{\k+\q,\tau}c_{\k,\tau'}\, .
\end{equation}
In this expression, both $\q$ and $\k$ lie in the mini-Brillouin zone, and $\g$ is a moir\'e reciprocal lattice vector. The operators $c_{\k,\tau} = (c_{\k,\tau,\uparrow}, c_{\k,\tau,\downarrow})^T$ annihilate an electron with momentum $\k$ in the mini-band of valley $\tau$. Note that since we are only considering one band per valley, we can use the valley index $\tau$ to label the mini-bands. The form factors are defined using the moir\'e Hamiltonian Bloch states $|u_\tau(\k)\rangle$ as
\begin{equation}
\lambda_{l,\g}^{\tau,\tau'}(\q,\k) = \langle u_{\tau}(\k+\q)|S_{\g}P_l|u_{\tau'}(\k)\rangle\,,
\end{equation}
where $P_l$ projects onto layer $l$ and $S_{\g}$ is a matrix with entrees $[S_\g]_{\g_i,\g_j} = \delta_{\g_i,\g+\g_j}$, where $\g,\g_i$ and $\g_j$ are moir\'e reciprocal lattice vectors. Using the projected density operators, we write the Coulomb Hamiltonian as the sum of an intra-valley and an inter-valley parts

\begin{eqnarray}
\tilde{H}_C & = & H_{V} + H_{IV}\, ,
\end{eqnarray}
where $H_V$ is U$(2)_+\times$U$(2)_-$ symmetric. Here we are only interested in the inter-valley part, which takes the form

\begin{equation}\label{IV}
H_{IV} =  \frac{1}{2NA}\sum_{\q,\k,\k'} \sum_{\tau}V^{C}_{\tau}(\q,\k,\k')c^\dagger_{\k+\q,-\tau}c_{\k,\tau}c^\dagger_{\k'-\q,\tau}c_{\k',-\tau}\ ,
\end{equation}
where the flat-band projected interaction potential, defined to include the form factors, is given by
\begin{eqnarray}
V^{C}_{\tau}(\q,\k,\k') & = &  \sum_{l,l',\g}V_{ll'}(\q+\g+2\X) \times \\
 & & \lambda^{\tau,-\tau}_{l,\g}(\q,\k)\lambda^{-\tau,\tau}_{l',-\g}(-\q,\k')\nonumber
\end{eqnarray}
In the above expression, we use $\X$ to denote the position of the center of the mini-Brillouin zone at the monolayer $K$ valleys (see Appendix \ref{app:MH} for additional details). Using a standard Fierz identity we can write $H_{IV}$ as the sum of an inter-valley density-density interaction and an inter-valley Heisenberg or Hund's coupling \cite{YaHuiZhang1}. We focus only on the SU$(2)_+\times$SU$(2)_-$ breaking term, i.e. the inter-valley Heisenberg term. From Eq. \eqref{IV} we see that it is of the form

\begin{eqnarray}\label{CJ}
H_{C,J} & = & -\frac{1}{NA}\sum_{\q,\k,\k'} \sum_{\tau}V^{C}_{\tau}(\q+\k'-\k,\k,\k') \\
& & \times \sum_i \left(c^\dagger_{\k'+\q,-\tau}\frac{s^i}{2}c_{\k',-\tau}\right)  \left(c^\dagger_{\k-\q,\tau}\frac{s^i}{2}c_{\k,\tau}\right)\nonumber \, ,
\end{eqnarray} 
where $s^i$ are the Pauli matrices acting on spin indices. To see whether the Hamiltonian in Eq. \eqref{CJ} prefers ferro- or anti-ferromagnetically aligned spins in different valleys, we define the four Slater determinants $|\tau,s\rangle = (N_M!)^{-1/2}\prod_{\k}c^\dagger_{\k,\tau,s}|0\rangle$, where $N_M$ is the number of moir\'e unit cells. The relevant matrix element determining the inter-valley spin splitting in first order perturbation theory is given in terms of these Slater determinants as

\begin{align}
& \langle +,\uparrow;-,\uparrow|H_{C,J}|+,\uparrow;-,\uparrow\rangle \nonumber \\
& = -\frac{1}{4NA}\sum_{\k,\k'}\sum_{\tau} V^C_\tau(\k'-\k,\k,\k')
\end{align}
We have calculated this matrix element numerically, and found that to a very good approximation it can be written as a function of the inter-layer distance $d$ as

\begin{align}
\label{eq:IVCoulomb}
\frac{1}{N_M}\langle +,\uparrow;-,\uparrow|H_{C,J}|+,\uparrow;-,\uparrow\rangle \nonumber \\
\approx -(0.20 - 0.16\, e^{-\frac{4\pi}{3}\frac{d}{a}}) \text{ meV}\, ,
\end{align}
So the inter-valley Heisenberg coupling arising from Coulomb interaction is ferromagnetic, and its magnitude increases as a function of the inter-layer distance. This is a consequence of the phase structure of the flat band wave functions, which leads to the minus sign in front of the exponential factor.

Next to the Coulomb interaction, there is a second source of SU$(2)_+\times$SU$(2)_-$ symmetry breaking, which comes from lattice-scale phonons near the $K$ points of the graphene Brillouin zone. As discussed in detail in Appendix \ref{app:elPh}, the phonon-induced inter-valley coupling projected into the flat bands is
\begin{eqnarray}\label{PH1}
H_{PH} & = & -\frac{g_{ph}}{N}\sum_{\q,\k,\k'}\sum_{\tau}V^{PH}_{\tau}(\q,\k,\k') \\
 & & \times c^\dagger_{\k+\q,-\tau}c_{\k,\tau}c^\dagger_{\k'-\q,\tau}c_{\k',-\tau}\,, \nonumber
\end{eqnarray}
where the phonon interaction strength is approximately $g_{ph}\approx 630$ meV. The phonon mediated interaction potential is expressed in terms of the form factors $f^\tau_{l,\g}(\q,\k)=\langle u_{-\tau}(\k+\q)|\sigma^xS_\g P_l|u_\tau(\k)\rangle$ as

\begin{equation}
V^{PH}_{\tau}(\q,\k,\k') = \sum_{l,\g}f^\tau_{l,\g}(\q,\k)f^{-\tau}_{l,-\g}(-\q,\k')\, .
\end{equation}
As before, we can use a Fierz identity to isolate the SU$(2)_+\times$SU$(2)_-$ symmetry breaking part of the Hamiltonian in Eq. \eqref{PH1}. We find 

\begin{eqnarray}\label{PHJ}
H_{PH,J} & = & \frac{2g_{ph}}{N}\sum_{\q,\k,\k'} \sum_{\tau}V^{PH}_{\tau}(\q+\k'-\k,\k,\k') \\
& & \times \sum_i \left(c^\dagger_{\k'+\q,-\tau}\frac{s^i}{2}c_{\k',-\tau}\right)  \left(c^\dagger_{\k-\q,\tau}\frac{s^i}{2}c_{\k,\tau}\right)\nonumber \, ,
\end{eqnarray} 
The relevant matrix element for the phonon induced inter-valley coupling Hamiltonian is

\begin{align}
& \langle +,\uparrow;-,\uparrow|H_{PH,J}|+,\uparrow;-,\uparrow\rangle \nonumber \\
&  = \frac{g_{ph}}{2N}\sum_{\k,\k'}\sum_\tau  V^{PH}_{\tau}(\k'-\k,\k,\k')
\end{align}
Evaluating this matrix element numerically, we find

\begin{align}
\frac{1}{N_M}\langle +,\uparrow;-,\uparrow|H_{PH,J}|+,\uparrow;-,\uparrow\rangle \approx 0.075 \text{ meV}
\end{align}
We see that the phonon induced inter-valley Heisenberg coupling is anti-ferromagnetic. Note that it is a significant fraction of the Coulomb inter-valley Heisenberg coupling in Eq.~(\ref{eq:IVCoulomb}) for $d\approx a$, so it cannot be neglected. In fact, if one would not take a finite layer separation into account in the Coulomb potential, the phonon contribution would dominate. We conclude that although the system at $\nu=2$ will most likely be ferromagnetic (FM) and spontaneously break time-reversal symmetry, we can not rule out the spin-valley locked state (SVL) where the electron spins are anti-aligned in different valleys ($\langle \tau^z \mathbf{s}\rangle \neq 0$). The ferromagnetic state with order parameter $\s$ was also recently found to describe the $\nu=2$ insulator observed in twisted double bilayer-graphene \cite{LiuHao,LeeKhalaf,CaoRodan,Shen}. The possibility of magnetic order in magic-angle tBLG was also previously discussed in Refs. \cite{Arraga,Thomson,KangVafek,Xie,Seo,WuKeselman,Wolf,Schrade19,Alavirad19}.

\section{Charged skyrmion excitations}

As mentioned previously, a skyrmion texture described by a unit vector field $\n(\r)$ in a spin polarized Chern band carries electric charge, as follows from the following general relation between the excess charge density $\rho(\r)$ and Pontryagin density \cite{Sondhi}:

\begin{equation}\label{skyrmcharge}
\rho(\r) = -\frac{C}{4\pi} \n(\r)\cdot(\partial_x \n(\r)\times\partial_y \n(\r))\, ,
\end{equation}
where $C$ is the Chern number. In order to identify skyrmions as the dominant charge carriers, we have to study their energetics, which is what we turn to next.

\subsection{Skyrmion energy with SU$(2)_+\times$SU$(2)_-$ symmetry}
For temperatures larger than the inter-valley Heisenberg coupling ($T \gtrsim 1$K), the spins from opposite valleys are decoupled via thermal fluctuations, while they remain ferromagnetically correlated within each valley due to the large Coulomb scale (as exemplified by the spin stiffness $\rho_s$ calculated below). Therefore, let us first ignore the inter-valley Heisenberg coupling and assume that the Hamiltonian is SU$(2)_+\times$SU$(2)_-$ symmetric. In that case, the lowest-energy skyrmions are skyrmions with topological charge $\pm 1$ in a single valley. Because the flat bands have Chern number $\pm 1$, these skyrmions have electric charge $\pm 1$ according to Eq. \eqref{skyrmcharge}. The energy of such a skyrmion is given by $E_{sk}=4\pi \rho_s$ \cite{BP75}, where $\rho_s$ is the spin stiffness. In Refs. \cite{Sondhi,Moon}, a mean-field expression for the spin stiffness of a spin polarized Landau level was derived. In Appendix \ref{app:rhos}, this expression is generalized to the case of electrons interacting via a density-density term of the form $\sum_\k \tilde{V}(\k) \rho(\k)\rho(-\k)$, projected onto a flat band with Berry curvature $\mathcal{F}(\k)$. Using the same approach as Ref. \cite{Moon}, we find the following approximate expression for the spin stiffness:

\begin{equation}\label{stiffness}
\rho_s = \frac{1}{8A}\left(\frac{1}{N}\sum_{\k'}\mathcal{F}(\k')^2\right) \left(\frac{1}{N}\sum_\k \tilde{V}(\k)f^2(\k)|\k|^2 \right)\, ,
\end{equation}
where $A$ is the area of the unit cell, $N$ is the number of unit cells and $f(\k) = |\lambda(\k,\k_0)|$ for some representative $\k_0$. The only approximation used to derive Eq. \eqref{stiffness} is that the magnitude of the form factor $|\lambda(\k,\q)|$ is independent of $\q$. If the Berry curvature is completely uniform throughout the Brillouin zone, Eq. \eqref{stiffness} reduces to the previously derived expression for Landau levels \cite{Sondhi,Moon}. From Eq. \eqref{stiffness}, we see that a non-homogeneous Berry curvature leads to a higher spin stiffness, and therefore a higher skyrmion energy.

If we apply Eq. \eqref{stiffness} to tBLG, we find
\begin{eqnarray}
\rho_s & \approx & \frac{1}{8A_M}\left(\frac{1}{N_M}\sum_{\k'\in \text{mBZ}}\mathcal{F}(\k')^2\right)\times \nonumber\\
 & & \left(\frac{1}{N}\sum_{\k\in\text{mBZ}}\sum_{\g} V(\k+\g)f_\g^2(\k)|\k|^2 \right)\, ,
\end{eqnarray}
where $A_M$ is the area of the moir\'e unit cell, $N$ is the number of mono-layer graphene unit cells, $N_M$ the number of moir\'e unit cells, $\g$ again denotes the moir\'e reciprocal lattice vectors, $V(\k)$ is the screened Coulomb potential defined in Eqs. \eqref{eq:ScrC} and \eqref{eq:dD} (with $d=0$), and $f_\g(\k)=|\sum_ l \lambda^{++}_{l,\g}(\k,\K_+/2)|$ (recall that $\K_+$ is the mini-BZ $K$-point). The reason for defining $f_\g(\k)$ with respect to the momentum point $\K_+/2$ instead of the $\Gamma$ point is that we found $|\sum_ l \lambda^{++}_{l,\g}(\k,\q)|$ to be largely independent of $\q$, \emph{except} near $\Gamma$.

The energy cost of a well-separated skyrmion pair $E_{2,sk}=8\pi\rho_s$ is to be compared with the energy cost of a particle-hole excitation in the spin polarized flat band. Using the same approximation as for the calculation of $\rho_s$, this energy cost in a mean-field decoupled Hamiltonian is readily found to be

\begin{eqnarray}\label{PH}
E_{ph}& = & \frac{2}{N}\sum_{\k,\g}V(\k+\g)f^2_\g(\k)\, ,
\end{eqnarray}
which agrees with the expression for the energy of a well-separated particle-hole pair in the spin polarized lowest Landau level \cite{Girvin2}, and sets the scale for the mean-field gap $M_S$.

\begin{center}
\begin{figure}
\includegraphics[scale=0.3]{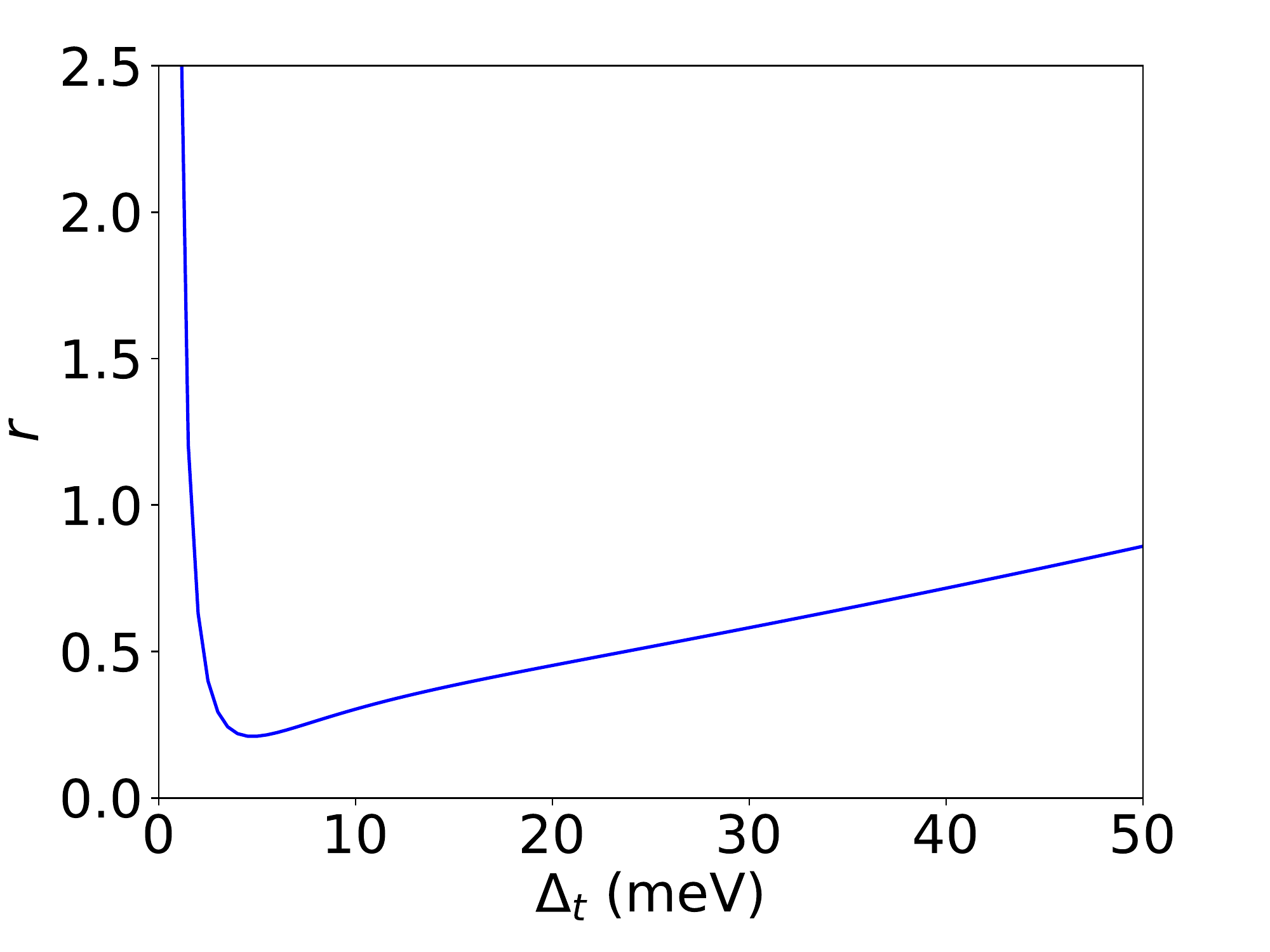}
\caption{Ratio $r=E_{2,sk}/E_{ph}$ of the energy of a well-separated skyrmion pair over the energy of a particle-hole excitation in a SU$(2)_+\times$SU$(2)_-$ symmetric model as a function of the sublattice splitting on the top graphene layer.}\label{fig:ratio}
\end{figure}
\end{center}

In Fig.~\ref{fig:ratio} we plot the ratio $r=E_{2,sk}/E_{ph}$ of the energy of a skyrmion pair over the energy of a well-separated particle-hole pair for a dual-gate screened Coulomb potential, as a function of the sublattice splitting $\Delta_t$ on the top layer. The shape of this curve is completely determined by the distribution of the Berry curvature over the mini-Brillouin zone. From Fig.~\ref{fig:ratio} we see that $r$  initially decreases very quickly, until it reaches a minimum at $\Delta_t\approx 5$ meV. This decrease follows from the fact that the Berry curvature is initially peaked at the $K$ points because of the Dirac cones in the $\Delta_t=0$ band spectrum, but starts to smoothen out when $\Delta_t$ increases. The ratio $r$ reaches a minimum for $\Delta_t\approx 5$ meV. After this minimum, the Berry curvature starts to accumulate again, this time at the $\Gamma$ point. Now the spin stiffness increases only slowly with $\Delta_t$. This is because a large value of $\Delta_t$ is required in order to close the gap between the flat band and the higher dispersive band at $\Gamma$, at which point the Berry curvature would also become singular. However, for realistic values of $\Delta_t$, we see that the skyrmion-pair energy is around $40$ to $45$ percent of the particle-hole energy. For example, with $\Delta_t=15$ meV, we find a skyrmion-pair energy of $\approx 21$ meV, and a particle-hole energy of $\approx 48$ meV. 

The  energy of a skyrmion in a single valley will increase when the inter-valley Heisenberg coupling is taken into account because this term wants to lock the spin moments in both valleys together and therefore penalizes a skyrmion texture made from the spins in only one valley but not the other. In the next section, we study the effect of non-zero inter-valley Heisenberg coupling in more detail.

\subsection{Effective field theory description}
\label{ssec:EFT}
In this section, we compute the energy of a charge $e$ skyrmion in a single valley with non-zero inter-valley Heisenberg coupling. We take into account the change of the ground state due to the external magnetic field, but neglect the back-reaction of spins in the opposite valley in response to the formation of a single skyrmion. We expect this to be a good approximation in the regime where the inter-valley exchange, parameterized by $\bro$, is weaker than the spin stiffness $\rho_s$ in each individual valley; this is the case for tBLG on HBN as shown by our numerical estimates ($\bro/\rho_s \approx 0.1$).

First, we consider the ferromagnet. A single skyrmion in one valley contains spins which are not aligned with the spins in the other valley, and also with the external magnetic field $B_\perp$ which aligns all spins with itself in the ground state. The core-size (and energy) of a skyrmion is determined by the competition between the Coulomb repulsion and exchange energy loss due to decoupling with spins from the opposite valley (determined by $\bro$), and with $B_\perp$. For small $\bro$ and $B_\perp$, the skyrmion would be large as it would try to minimize Coulomb repulsion. On increasing $B_\perp$, the Zeeman energy dominates and the skyrmion size saturates to a small value of the order of moir\'e lattice spacing $a_M$. In this limit, the skyrmion energy also saturates to a maximum value; and a skyrmion-antiskyrmion pair resembles a particle-hole pair.

\begin{figure*}[t]
\centering 
\begin{subfigure}[t]{0.6\textwidth}
\vspace{-4.2 cm}
\includegraphics[width=\textwidth]{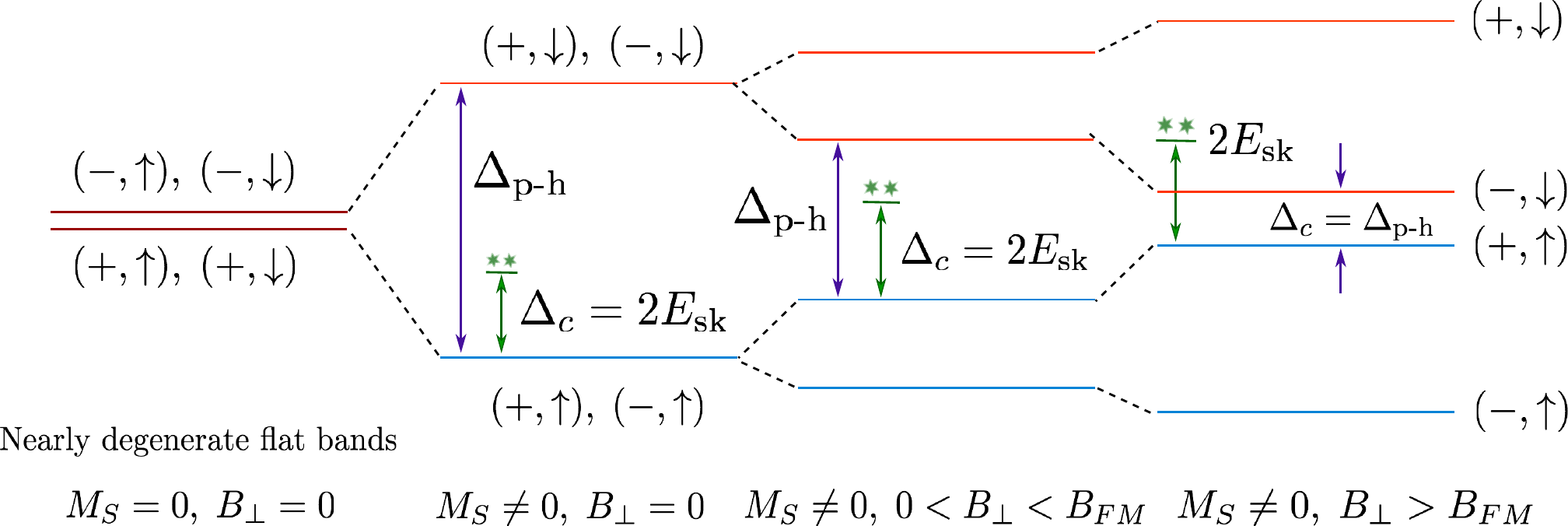}\vspace{0.63 cm}\caption{}
\end{subfigure}\hspace{0.4 cm}
\begin{subfigure}[t]{0.35\textwidth}
\includegraphics[width=\textwidth]{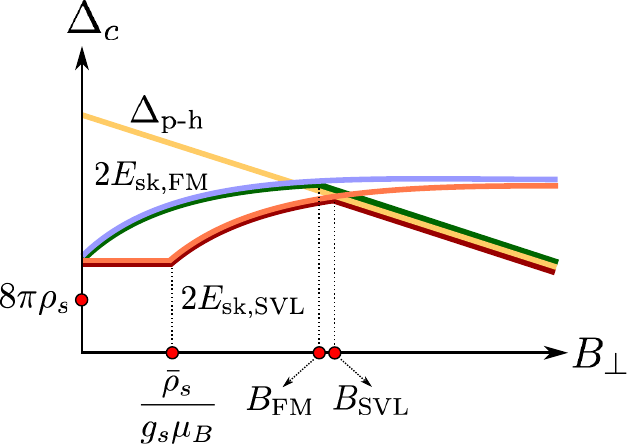}\caption{}
\end{subfigure}\hspace{0.4 cm}
\caption{(\textbf{a}) Band-splitting and skyrmion gap in the ferromagnet as a function of $B_\perp$. (\textbf{b}) Green (brown) line schematically depicts the charge gap $\Delta_c$ as a function of $B_\perp$ for the ferromagnet (spin-valley locked state). $\Delta_c$ increases till $B_\perp = B_{FM}$ ($B_{SVL}$) when the energies of the skyrmion-pair and particle-hole pair cross, and then drops. } \label{fig:bandSplittings}
\end{figure*}

To illustrate this schematically, we make an (over-simplified) estimate the energy $E_{sk}$ of a two-dimensional skyrmion of linear size $R$, which is given by the sum of its elastic energy $E_{\text{el}}$, Coulomb energy $E_C$ (for simplicity we temporarily ignore screening) and Zeeman-energy $E_Z$ that receives contribution from both the inter-valley coupling and the external magnetic field $B_\perp$:
\beq
 E_{sk} &\approx& 4 \pi \rho_s + \frac{e^2}{4\pi \epsilon R} + \left(g_s \mu_B B_\perp + \frac{\bro}{2} \right)\left(\frac{R}{a_M}\right)^2 \implies \nn R_{\text{opt}} &\approx& \left( \frac{e^2 a_M^2}{4\pi \epsilon  (g_s \mu_B B_\perp + \bro/2)} \right)^{1/3} \equiv \left( \frac{a_M^2 \ell^2_{\tilde{B}}}{a_0}\right)^{1/3}
\eeq
 where $a_0 = \frac{4\pi \epsilon }{m_e e^2}$ is the effective Bohr radius, $\ell_{\tilde{B}} = \sqrt{\hbar/e [B_\perp + \bro/(2 g_s \mu_B)]}$ is the effective magnetic length and $a_M$ is the moir\'e lengthscale. At the optimal length-scale, the energy of the skyrmion is given by 
 \beq
&& E_{sk}(B_\perp) - E_{sk}(B_\perp = 0) \approx \frac{e^2}{4\pi \epsilon a_M} \left( \frac{a_0 a_M}{\ell^2_{\tilde{B}}}\right)^{1/3} \nn && ~~~~~~~~~~~~~~~~~~~~~~ \propto \begin{cases} B_\perp, \text{ for } g_s \mu_B B_\perp \ll \bro \\ B_\perp^{1/3} \text{ for } \bro \ll g_s \mu_B B_\perp \end{cases}
\label{eq:EskApprox}
 \eeq
 Therefore, $E_{sk}$ first increases linearly, and subsequently sublinearly in $B_\perp$ for small $B_\perp$; this feature remains valid even in presence of screening and can contribute to an increasing charge gap on turning on $B_\perp$.

Next, we turn to a continuum field theory for a more accurate estimate of the skyrmion energy. The effective Lagrangian density for the ferromagnet can be described by the following two-component O(3) non-linear $\sigma$ model:
\beq
\mathcal{L} &=& \sum_{\tau=\pm} \bigg[ n S \bigg( \bm{A}[\n_\tau] \cdot \partial_t \n_\tau(\r)  + g_s \mu_B \B \cdot \n_\tau(\r) \bigg) \nn && ~~~~~~ - \frac{\rho_s}{2} (\nabla \n_\tau (\r))^2 \bigg] - \frac{nS^2 \bro}{2} [\n_+(\r) - \n_-(\r)]^2 \nn
&& ~~~~~~ - \frac{1}{2} \int d\rp \, V(\r - \rp) \rho(\r) \rho(\rp) 
 \label{eq:Leff}
 \eeq
where $\bm{A}[\n_\tau]$ corresponds to the vector potential of a unit monopole with $\nabla_\n \times \bm{A}[\n_\tau] = \n_\tau$, and $\n_\tau(\r)$ lies on the 2-sphere ($\n_\tau \cdot \n_\tau = 1$). $\rho(\r) = \sum_\tau \rho_\tau(\r)$ with $\rho_\tau(\r) = - \frac{C_\tau}{4\pi} \n_\tau \cdot (\partial_x \n_\tau \times \partial_y \n_\tau)$ is the topological charge density of the skyrmion ($C_\tau  = \mp 1$ for valleys labeled $\tau = \pm$), $\bro$ is the inter-valley spin-stiffness, $n = 2/(\sqrt{3}a_M^2)$ is the density of electrons and $S = 1/2$ is the electron spin ($\hbar = 1$). In the ground state, $\n_+(\r) = \n_-(\r)$ for the ferromagnet so the term with $\bro$ does not contribute. 

To calculate the energy of a skyrmion configuration, it is convenient to use complex coordinates $z = x + i y$, and write the single skyrmion texture in terms of a complex analytic function $W(z)$ as follows \cite{BP75}. 
\beq
n_x - i n_y = \frac{2 W(z)}{1 + |W(z)|^2}, \; n_z = \frac{1 - |W(z)|^2}{1 + |W(z)|^2}
\label{eq:Wans}
\eeq
As shown in Appendix \ref{app:EnSky}, we find that the field theory yields the following energy for the skyrmion ansatz $W(z) = R/z$ after optimizing its size $R$ ($\alpha$ is an O(1) numerical constant).
\beq
E_{sk} = 4 \pi \rho_s + \alpha E_C \left[ \left( \frac{\Delta}{E_C} \right) \ln \left( 1+ \frac{E_C}{\Delta} \right) \right]^{\nu}
\label{eq:EsAna}
\eeq
where $\nu = 1/2$ (1/3) for strongly gate-screened (unscreened) Coulomb interaction (see Eq.~(\ref{eq:ScrC})), and $\Delta = g_s \mu_B B_\perp + \bro/2$ is the effective Zeeman energy-scale in a given valley. We conclude that irrespective of the precise details of screening, $E_{sk}$ increases sub-linearly with $B_\perp$ for small external fields. Though this effective theory cannot capture large $B_\perp$ when lattice-scale effects become important, the skyrmion energy is expected to saturate as a skyrmion pair gets squeezed to a particle-hole pair. 

\begin{figure*}[t]
    \centering
    \includegraphics[scale=0.8]{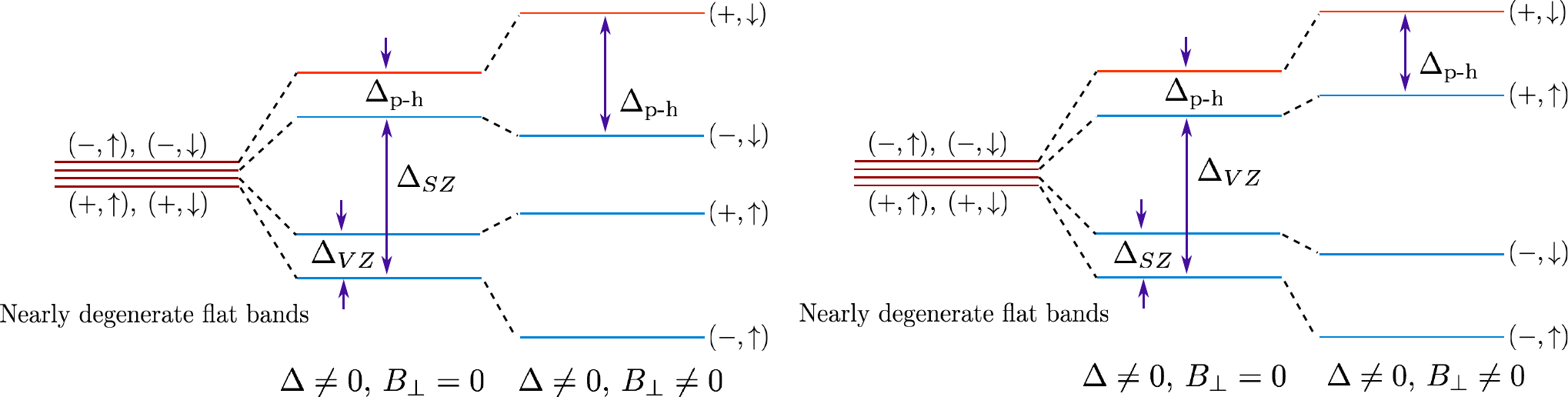}
    \caption{The band-gap evolution at $\nu = 3$ as a function of $B_\perp$, for $\Delta_{VZ} < \Delta_{SZ}$ and $\Delta_{VZ} > \Delta_{SZ}$. We have assumed that $g_v < 0$, $g_s > 0$ and $|g_v| > |g_s|$, following Refs.~\cite{AHpaper,YaHuiZhang3}.}
    \label{fig:bandSplittingsNu3}
\end{figure*}

We can estimate the energy and size of a skyrmion for screened Coulomb coupling with a screening length of the order of $a_M$. Taking $E_c \approx \rho_s \approx 1$ meV, we find that the correction to the elastic energy of the skyrmion is $\approx 1$ meV for $B_\perp = 0$ and $\bro =  0.12$ meV. This implies that the skyrmion-antiskyrmion pair still costs lower energy than the particle-hole pair. The net magnetic moment carried by the skyrmion is approximately $2.7 \, g_s \mu_B$, so we are in the regime where the skyrmion size is quite small. Hence, the exact numerical estimates from our continuum model are not likely to be very accurate; however, they are robust to small microscopic deformations of the Hamiltonian and provide a reasonable sense of the relevant energy scales. 

For the spin-valley locked state, we replace $\n_- \rightarrow -\n_-$ in Eq.~(\ref{eq:Leff}). While the Zeeman gap $\Delta = \bro/2$ is identical to the ferromagnet for $B_\perp = 0$, turning on $B_\perp$ causes spins from opposite valleys to cant towards itself, changing the ground state (however, spins within one valley remain ferromagnetically aligned). Interestingly, the effective Zeeman gap for a single valley ($\Delta$) remains constant until the field reaches the critical value $B_\perp = \bro/(g_s \mu_B)$, at which point a transition to the ferromagnetic state occurs (see Appendix \ref{app:EnSky}). Once the system is ferromagnetic, the skyrmion energy increases linearly as discussed above. To summarize, we find the following behavior for $\Delta$:
\beq
\Delta = g_s \mu_B \tilde{B} = \begin{cases} \frac{\bro}{2}, \;  B_\perp < \frac{ \bro}{g_s \mu_B} \\  g_s \mu_B B_\perp - \frac{\bro}{2},\; B_\perp \geq \frac{\bro}{g_s \mu_B} \end{cases}
\label{eq:BeffAF}
\eeq
Accordingly, the skyrmion size also remains fixed till $B_\perp = \bro/(g_s \mu_B)$ and then gradually decreases as $B_\perp$ is tuned up further.

\subsection{Charge gap in a magnetic field at $\nu = 2$}

Having established that a skyrmion is the lowest energy charge $e$ excitation for small external fields, we now turn to the longitudinal resistivity $\rho_{xx}$ as a function of $B_\perp$. We assume that the insulator at $\nu = 2$ has an activated $\rho_{xx}$ which is governed by the gap $\Delta_c$ to the charged excitation that costs the lowest energy. Because of the valley- and spin-Zeeman terms in Eq.~(\ref{ham1}), the bandgap decreases with increasing $B_\perp$, and hence the gap to exciting an electron to an empty band decreases. On the other hand, the single charge $e$ skyrmion gap for the ferromagnet increases as $B_\perp^\nu$ ($1/3 \leq \nu \leq 1/2$ depending on the nature of screening) for small fields $B_\perp$. Therefore, the overall charge gap $\Delta_c = \text{min}\{2 E_{sk}(B_\perp), 2|M_S| - \mu_B |g_{v,max} B_\perp| \}$ will initially increase as a function of $B_\perp$, and then start dropping when the valley-Zeeman term dominates, as schematically depicted in Fig.~\ref{fig:bandSplittings}. Assuming that the behavior of the resistivity is determined entirely by the activation gap $\Delta_c$, charge $e$ skyrmions can explain the peculiar behavior of $\rho_{xx}(B_\perp)$ \cite{Goldhaber}. For the spin-valley locked state, the gap remains constant till $B_\perp \approx \bro/(g_s \mu_B) $, and then increases; therefore it appears unlikely that the ground state is the spin-valley locked state based on the transport data. This agrees with the results of Section \ref{symmbreaking}, where we found the net inter-valley Heisenberg coupling to be ferromagnetic.

\subsection{Charge gap at $\nu = 3$}

Next, we turn our attention to the $\nu = 3$ state and discuss predictions for the charge gap in presence of $B_\perp$, assuming it is insulating in a high-quality sample. The anomalous Hall effect and evidence of edge transport \cite{Goldhaber} can be explained by a single spin and valley polarized hole-band. Equivalently, three of the four conduction bands are filled; for concreteness let us assume these are $(\tau^z,s^z)=(+,\uparrow), (-,\uparrow)$ and $(+,\downarrow)$. If the lowest energy charged excitations are skyrmions, then the energy of a single isolated skyrmion is be given by Eq.~(\ref{eq:EsAna}). In particular, the elastic energy $4\pi \rho_s$ of the skyrmion should remain unchanged as the spin-stiffness $\rho_s$ is insensitive to the valley or spin quantum number of the conduction band. The effective magnetic field seen by the skyrmion $\tilde{B}$ is given by the sum of the external field $B_\perp$ and the internal field which is proportional to $\bro$ and the internal Zeeman field from the ordered moments of the remaining filled bands. In our mean-field picture, the $(+,\uparrow)$ and $(+,\downarrow)$ states form a spin-singlet at each $\k$. Therefore, skyrmions cannot be excited in the $\tau = +$ valley. A skyrmion excitation is possible in the $\tau = -$ valley, starting with electrons in the $(-,\downarrow)$ band. Such a skyrmion will \textit{see} no background ordered moment, and therefore have a lower energy $E_{sk}$ given by Eq.~(\ref{eq:EsAna}) with $\tilde{B} = B_\perp$. The charge gap $\Delta_c$ is just $2 E_{sk}$. 

At higher external fields, we expect the charge gap to be set by the particle-hole gap, as the skyrmion energy increases with $B_\perp$. Note that although the degeneracy between the four conduction bands is spontaneously broken at $B_\perp = 0$, turning on an infinitesimal $B_\perp$ automatically chooses an arrangement of the bands via the valley and spin Zeeman terms in the Hamiltonian. The behavior of the particle-hole gap as a function of $B_\perp$ depends on the sequence in which these bands are ordered with energy, which in turn depends on the interaction induced valley-Zeeman and spin-Zeeman gaps at zero $B_\perp$. If the valley-Zeeman gap is larger than the spin-Zeeman gap, the top two bands are spin-split and they move apart under an applied $B_\perp$ via the single-particle Zeeman shift with a constant g-factor $g_s = 2$. In contrast, if the spin-Zeeman gap dominates and the top two bands are valley-split, then the particle-hole gap increases with $B_\perp$, but with a g-factor $g_v(\k_0)$ where $\k_0$ corresponds to the point where the gap is minimal at $B_\perp = 0$. The situation is depicted schematically in Fig.~(\ref{fig:bandSplittingsNu3}), where we neglect the dispersion of the flat bands (which is justified for a spatially uniform order parameter, as the gap magnitude is set by the larger Coulomb scale). 

To summarize, at $\nu = 3$ the skyrmion energy is expected set the charge gap at $B_\perp = 0$, leading to a non-linear onset with $B_\perp$. At intermediate fields the skyrmion-antiskyrmion energy will exceed the particle-hole energy. In this regime gap will continue to increase with $B_\perp$, but linearly. Further, the coefficient of linear increase can tell whether the valley Zeeman gap is larger than the spin-Zeeman gap at $B_\perp = 0$, or vice-versa. At even larger values of $B_\perp$, the valence bands which we have neglected till now may come close the Fermi level, resulting in a decrease of the charge gap. 

\section{Skyrmion pairing}
In this section, we show that charge $e$ fermionic skyrmions have a generic tendency to attract and bind into charge $2e$ bosonic pairs, leading to the exotic possibility of quantum phases (like superconductivity) that can arise from skyrmion pairing at finite density at $T = 0$. Although we use a semiclassical description for the energetics to maintain an analytical handle and pinpoint the physical mechanism of pairing, the small size of the skyrmions for parameters relevant to tBLG (see section \ref{ssec:EFT}) motivates us to consider these skyrmions (or skyrmion-pairs) as charged quantum quasiparticles. Therefore, we can envision transitions to quantum liquid (superconductor) or quantum solid (Wigner crystal) phases of $2e$ skyrmion-pairs at small but finite density of charge carriers, in the same spirit as band structures or phase transitions of charge-neutral quantum skyrmions have been considered in two-dimensional chiral magnets \cite{Balents2016}.  

We first consider two skyrmions from the same Chern band (same valley). If they have opposite phases in the plane normal to the spin-ordering axis (x-y plane in our scenario), they will always attract at large distance scales (an opposite phase-winding skyrmion can be obtained by $\n = (n_x,n_y,n_z) \rightarrow (-n_x,-n_y,n_z)$ and has the same topological and electric charge). The physical reason is simple: for a pair of well-separated skyrmions of opposite phases (the distance between their centers $2L$ is much larger than the typical skyrmion size $R$, but smaller than the spin-correlation length $\xi_s$), the components of the spin pointing normal to the effective field $\tilde{B}$ are quenched at distance $L \ll \xi_s$. This lowers the effective Zeeman energy, which is present in tBLG due to intervalley coupling even at $B_\perp = 0$. Indeed, we show below that the effective Zeeman energy gain is logarithmic and this results in a $L^{-1}$ attractive force between these skyrmions that always prevails the $L^{-2}$ Coulomb replusion at large distances \cite{Sondhi2,NK_PRL98} (or a screened Coulomb repulsion, which decays exponentially at distances larger than the screening length). Therefore the skyrmions prefer to be paired at the lowest energy scales (akin to vortices in a U(1) superfluid below the Berezinskii-Kosterlitz-Thouless transition temperature $T_{BKT}$). 

\begin{figure}
    \centering
    \includegraphics[scale=0.7]{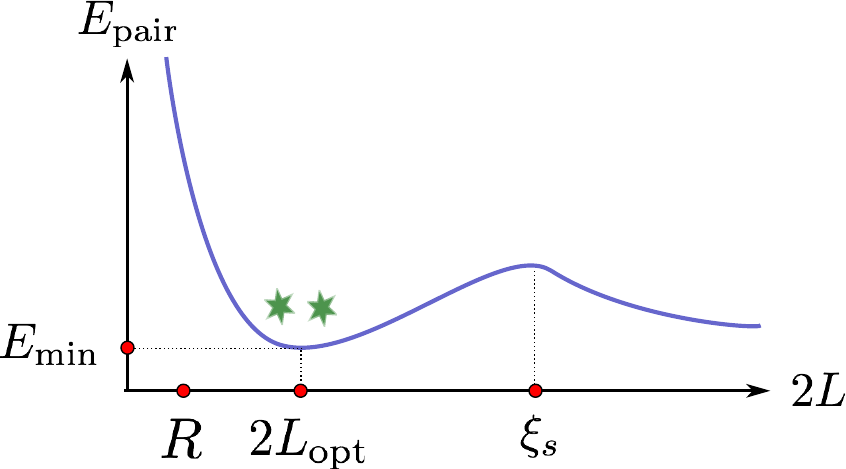}
    \caption{Schematic depiction of the skyrmion-pair potential from a single valley. The skyrmions can form a bound state if the minima at $2L_\text{opt}$ is deeper than minima at $L \rightarrow \infty$.}
    \label{fig:SkyEnPair}
\end{figure}

We now consider an opposite phase skyrmion pair configuration in the ferromagnet, with a distance $2L$ between their centers.
\beq
W(z) = \frac{R}{z-L} - \frac{R}{z + L}
\eeq
The energy of the skyrmion-pair $E_{\text{pair}}$ can be computed using the effective field theory in Eq.~(\ref{eq:Leff}); the details are relegated to Appendix \ref{app:EnSkyPair}. 
\beq
E_{\text{pair}} &=& E^{\text{el}}_{\text{pair}} + E^Z_{\text{pair}} + E^\text{C}_{\text{pair}} \nn &=& 8 \pi \rho_s + \frac{8 \pi g_s \mu_B \tilde{B} R^2}{\sqrt{3} a_M^2} \ln \left( \frac{2L}{R} \right) + \frac{e^2}{4 \pi \epsilon (2L)} \nn
\label{eq:ESkPair}
\eeq
where $ \tilde{B} =  B_\perp + \frac{\bro}{2 g_s \mu_B}$ is the effective Zeeman field at $\nu = 2$. Therefore, we confirm that the skyrmion-pair attracts at distances $L$ larger than $R$ but smaller than $\xi_s$, as depicted schematically in Fig.~\ref{fig:SkyEnPair}. 

At finite density the charge $2e$ bosonic skyrmion-pairs can either condense to form a superconductor, or form into a Wigner crystal phase to minimize Coulomb repulsion \cite{NK_PRL98}. Such superconductivity may be aided by gate-screening of Coulomb interaction, or suppressed by the Magnus force felt by a skyrmion-pair from the same Chern band \cite{Stone96}. The exact phase diagram requires an involved study we will not attempt here; instead we focus on the symmetry properties of superconductor obtained by such a condensate. 

To derive the quantum numbers of the skyrmions and skyrmion-pairs, we follow the approach in Ref.~\onlinecite{NK_PRL98}. we note that if we write $n_x + i n_y \sim e^{i \phi} \sin \theta $, the classical phase space variables $(\phi, n_z)$ can be promoted to canonically conjugate quantum operators $(\hat{\phi}, 2\hat{s}_z)$. The infinite degeneracy with respect of rotation of phase $\phi$ for a single classical skyrmion texture translates to a fixed quantum number for the total spin $S_z = (n/2)\int d\r [n_z(\r)-1]$ (recall that $n$ is the electron density, and we have subtracted out the background spin from the ground state). The size of the quantum skyrmion $R$ takes the closest possible value to the classical minimum to ensure a half-integer spin $S_z$. The same should be true for a skyrmion-pair, the quantum $2e$ boson corresponding to  classical texture $\n_{\text{pair}}(\r)$ carries a quantized integer spin given by 
\beq
S_z = \frac{n}{2} \int d\r \, [\n_{\text{pair}}(\r) \cdot \hat{z} - 1]
\eeq
The $2e$ bosonic pair carries non-zero net spin; therefore its condensation breaks time-reversal, and ferromagnetism persists into the superconductor. This is easiest to see in the small-size limit when the skyrmion pair resembles a hole pair, it carries spin $S=1$, and its condensation leads to triplet superconductivity. However, the symmetry properties of the superconductor do not change away from this limit, where our proposed mechanism is operative. Hence, the quantum phase transition (QPT) from the ferromagnet to the superconductor only breaks U(1) charge conservation, and is described by the Abelian Higgs model (charged $2e$ scalar coupled to a U(1) electromagnetic gauge field). Further, since the $2e$ boson carries charge $\pm 2$ under U(1)$_v$ (depending on the valley/Chern sector), it transforms non-trivially under moir\'e lattice translations and its condensation will lead to broken translation symmetry with a three-fold enlarged unit cell. A similar scenario holds for the spin-valley locked state as well, making appropriate modifications to $\tilde{B}$ using Eq.~(\ref{eq:BeffAF}).

Next, we consider skyrmion pairing from opposite Chern sectors (or equivalently, from opposite valleys). In case of the ferromagnet, a charge 2e pair requires pairing a skyrmion from one valley with an antiskyrmion from the opposite valleys (since they have opposite Chern numbers). This does not lead to any effective Zeeman energy gain, and is therefore not favorable. However, skyrmion pairing of opposite charges from opposite valleys is favored by both Coulomb and effective Zeeman terms (as it locally preserves the inter-valley ferromagnetic configuration when the skyrmions sit on top of each other with $\n_+(\r) = \n_-(\r)$). Such a skyrmion pair again carries a large spin. The resulting inter-valley coherent state breaks valley U(1) and spin-rotation, and the QPT is described by a complex scalar field theory. Note that this state is distinct from the conventional time-reversal preserving IVC phase discussed in the context of tBLG \cite{YaHuiZhang3}. Regardless, a uniform condensate of such skyrmion pairs is also precluded by the opposite Chern number of the bands. To understand this, one can again resort the small size (or large field) limit, when the skyrmion-antiskyrmion pair reduces to a particle-hole or exciton pair carrying a net spin $S = 1$. Since the argument of Ref.~\onlinecite{AHpaper} relies solely on topological considerations and is independent of the spin of charge carriers, we expect such a uniform condensate to be energetically unfavorable. A lattice of skyrmion-antiskyrmion pairs (analogous to the exciton-vortex lattice discussed in Ref.~\onlinecite{AHpaper}) offers an attractive alternative, but more detailed investigations are required to establish its stability. 

For the spin-valley locked state, a skyrmion-antiskyrmion pair from opposite valleys (both with same charge) can avoid losing any exchange energy at zero $B_\perp$ by keeping spins locally anti-aligned ($\n_+(\r) = - \n_-(\r) \equiv \n(\r)$), and simultaneously quench the Coulomb energy cost by having a very large radius $R$ which is fixed by small anisotropies beyond the SU(2)$_+ \times$ SU(2)$_-$ symmetric limit. Such a charge $2e$ pair therefore only costs an elastic energy of $E_{\text{pair}} \approx 8 \pi \rho_s$. In analogy with the previous discussion, the quantum number $Q^a$ of the skyrmion-antiskyrmion pair under a generator $T^a \in \{ \s,\tau^z \s \}$ of the symmetry group SU(2)$_+ \times$ SU(2)$_-$ are given by:
\beq
Q^a = \frac{n}{2} \int d\r \, (\n_{sk}(\r) - \n_0)\cdot\Tr[(\tau^z \s) T^a]
\label{eq:SVLPairQNo}
\eeq
where $\n_{sk}(\r)$ is the skyrmionic texture in $\n(\r)$ and $\n_0 = (0,0,1)$ is the ground state configuration. From Eq.~(\ref{eq:SVLPairQNo}), we note that the superconductor formed by skyrmion pairing from opposite valleys in the spin-valley locked state preserves global spin-rotation symmetries, i.e, $Q^a = 0 ~\forall~ T^a \in \{\s\}$. Further, it also preserves time-reversal and translation (being neutral under U(1)$_v$). This necessarily implies that in case of a direct transition, the critical point that describes the QPT from the spin-valley locked state (breaks spin-rotation symmetry, preserves U(1) charge conservation) to the superconductor (which breaks U(1) charge conservation but preserves spin-rotation) is a deconfined quantum critical point. The critical theory for this transition has been discussed using a five-component 'super-spin' order parameter in Ref.~\cite{GS2008} that transforms as a vector under an emergent SO(5) symmetry. The defects of the spin-Hall like order parameter, which are skyrmion pairs, carry charge $2e$. Therefore proliferation of these defects leads to suppression of anti-ferromagnetic order and simultaneous appearance of superconductivity. Approaching from the opposite side, the defects of the superconductor, which are vortices, carry quantized spin. This can be seen via the critical theory with the Wess-Zumino-Witten term in Ref.~\cite{GS2008}; the latter endows a superconducting vortex with a spin-half. Hence, proliferation of vortices destroys superconductivity and simultaneously results in long-range magnetic order. 

Lastly, we note that if this mechanism is operative in tBLG, the critical temperature of the superconducting transition would be set by the Heisenberg coupling $J$ between the spins from opposite valleys (which provides the binding energy). From Eq.~(\ref{eq:IVCoulomb}), we therefore expect $T_c \sim J \sim 1 K$. An in-depth investigation of superconductivity via skyrmion-pairing, including a quantitative estimate of $T_c$ and a phase diagram as a function of doping, will be the subject of a forthcoming study \cite{KCBZV}.

\section{Discussion}

We have argued that the $\nu=2$ resistance peak observed in magic-angle tBLG aligned with hBN observed in Ref. \cite{Goldhaber,Serlin} arises from electrons filling a spin polarized band in each valley. The spins in different valleys are most likely aligned ferromagnetically, but we cannot completely exclude the possibility that there is anti-ferromagnetic alignment between the valleys. The precise nature of the inter-valley spin correlation depends on lattice-scale effects which determine the inter-valley Heisenberg coupling and are not accurately captured by our approach. However, irrespective of the spin alignment or anti-alignment between the valleys, we expect skyrmion excitations to be lower in energy than particle-hole excitations. Because of the the non-zero Chern number of the flat bands, these skyrmions carry charge $\pm e$, making them the most relevant charge carriers. Because skyrmions have a large effective $g$-factor, the spin-Zeeman term efficiently raises their energy, which we propose to be the origin of the increase in resistivity with out-of-plane magnetic field observed in Ref. \cite{Goldhaber} at $\nu=2$. We note that our diagnosis of a ferromagnetic insulator at $\nu = 2$ based on magnetotransport data is consistent with recent predictions of ferromagnetic insulating states at integer fillings of nearly flat bands based on exact diagonalization and DMRG studies of models appropriate to tBLG on hBN \cite{Cecile19}.

\textit{Experimental probes}: A natural question arises regarding experimental probes that distinguish between the different magnetic orders at $\nu = 2$, since  neutron-scattering experiments may be difficult due to the two-dimensional nature of the sample. The ferromagnet breaks time-reversal symmetry, and therefore can be probed using muon spin resonance. However, non-linear optical responses that are enhanced by orbital ferromagnetism in flat bands as suggested in Ref.~\cite{LiuDai19} will remain suppressed as there is no net valley-polarization at this filling. The spin-valley locked state breaks spin-rotation but preserves time-reversal (since opposite valleys carry opposite spin), and is comparatively harder to detect. We note that the collective magnons (which simultaneously involve both valleys) have different dispersions in the two cases (quadratic for FM, linear for spin-valley locked); further ferromagnetic magnons gap out under a magnetic field while antiferromagnetic magnons do not. Therefore, studying the magnetic contribution to specific heat or thermal conductivity; or performing spin-injection experiments (which can directly probe the magnon dispersion) at the sample-edge \cite{SCSS15,AFYoung2019} can distinguish these states. Since a skyrmion has a large number of flipped spins, one can sense a trapped skyrmion in an impurity potential via spin-polarized STM, or local magnetometers like a scanning nano-squid \cite{AFY2016} or a Nitrogen-Vacancy (NV) center \cite{Dovzhenko2018}. Finally, if the state is indeed an AFM, then applying a strong $B_\perp$ will cant the spins and change the ground state. As discussed, the charge $e$ skyrmion gap $\Delta_c(B_\perp)$ will behave very differently from the ferromagnet; it will stay constant till a critical field $B_c$ that induces a phase transition to FM. Hence, a careful study of the activation gap as a function of the magnetic field can distinguish these scenarios. The said phase transition to a FM and associated critical signatures may also be observed via thermodynamic probes. 

\textit{Outlook}: Recent theoretical and experimental works have shown that flat bands with non-zero Chern number are quite common in moir\'e materials \cite{YaHuiZhang1,LiuMa}. For example, Refs. \cite{Xie,Efetov,LiuKhalaf} found either from experiments or a self-consistent Hartree-Fock calculation that in certain regimes electron interactions in magic-angle tBLG unaligned with hBN lead to a spontaneous breaking of the $C_{2v}T$ symmetry protecting the Dirac cones, giving rise to mean-field bands with Chern number equal to either $\pm 2$ \cite{Xie,Efetov} or $\pm 1$ \cite{LiuKhalaf,Efetov}. In twisted double bilayer graphene the $C_2$ symmetry is broken explicitly on the single-particle level, and the flat bands have Chern number $2$ \cite{LeeKhalaf}. In Ref. \cite{ChenSharpe}, a Chern insulator at $\nu=1$ was observed in ABC trilayer-graphene on hBN, which can be understood from a Hartree-Fock study which predicts mean-field bands with Chern number $\pm 2$ at intermediate interaction strengths.

There is also mounting evidence that the insulating states at integer $\nu$ result from spontaneous symmetry breaking which lifts the spin and valley degeneracies, similar to what happens in quantum Hall ferromagnetism \cite{Sondhi,Girvin,Eisenstein}. The general picture that seems to emerge at present is that this spin and valley degeneracy lifting occurs in a valley-U$(1)$ preserving manner, i.e. without developing inter-valley coherence. For example, the anomalous Hall effect at $\nu=3$ in tBLG aligned with hBN observed in Refs. \cite{Goldhaber,Serlin} and the Chern insulator at $\nu=1$ in trilayer graphene \cite{ChenSharpe} can both naturally be attributed to a spontaneous valley polarization \cite{AHpaper,YaHuiZhang3,ChenSharpe}. The insulators at $\nu=1$ and $\nu=2$ observed in twisted double bilayer graphene in Ref. \cite{LiuHao,CaoRodan,Shen} were proposed to respectively be a valley-polarized and valley-singlet ferromagnet \cite{LeeKhalaf}. A priori, skyrmions could play a role in charge transport for any of these devices. However, this is less likely for bands with higher Chern numbers because the spin stiffness increases quadratically with $C$ \cite{Girvin,YaHuiZhang1}. We note that, interestingly, the $\nu=-2$ insulator observed in ABC stacked trilayer graphene on hBN \cite{ChenSharpe} also shows an increased resistance peak under an applied out-of-plane magnetic field. ABC stacked trilayer graphene has a large orbital $g$-factor \cite{YaHuiZhang2}, which means that the valley-Zeeman effect dominates the spin-Zeeman effect. Because of this, one expects that a slightly modified version of our discussion in the main text applies to this device as well. 

An important general open question concerns the connection between the insulators observed at integer fillings in moir\'e materials and the superconducting domes which result from doping these insulators. No superconducting domes were observed in Refs. \cite{Goldhaber,Serlin}, but this could be because the temperatures in these experiments were too high, or because of device quality. Further experimental studies are needed to either rule out superconductivity in magic-angle tBLG aligned with hBN, or to establish its existence and measure its response to different electric and magnetic fields. If superconductivity is observed, theory will have to come up with a pairing mechanism for the charge carriers which are doped into the insulator. In this work, we looked into the possibility of skyrmion pairing, but other mechanisms are possible of course. For example, Ref. \cite{LeeKhalaf} proposed a more conventional pairing mechanism driven by ferromagnetic spin fluctuations to explain the superconducting domes in twisted double bilayer graphene.

Finally, the precise connection between the insulators observed in magic-angle tBLG aligned with hBN, and those observed in the $C_{2v}$ symmetric devices \cite{Cao,Efetov} where the substrate does not significantly modify the single-particle physics, is not clear. Theoretically, one would like to understand what happens if one continuously turns off the hBN-induced sublattice splitting. It is likely that some insulators will undergo phase transitions, perhaps accompanied by changes in Chern number. Understanding this connection is an important missing piece in the moir\'e puzzle.

\subsection*{Acknowledgements} It is a pleasure to thank Zhen Bi, Rafael Fernandez, David Goldhaber-Gordon, Jiang Kang, Eslam Khalaf, Biao Lian, Hoi Chun Po, Louk Rademaker, Cecile Repellin, Todadri Senthil, Oskar Vafek, Ashvin Vishwanath, Fengcheng Wu, Andrea Young and Ya-Hui Zhang for stimulating discussions. SC is particularly thankful to Eslam Khalaf for clarifying the computation of skyrmion quantum numbers, and related collaborations. SC acknowledges support from the ERC synergy grant UQUAM via Ehud Altman. MZ and NB were supported by the DOE, office of Basic Energy Sciences under contract no. DE-AC02-05-CH11231. This work was finalized in part at the Aspen Center for Physics, which is supported by National Science Foundation grant PHY-1607611.

\bibliography{SkyrmionstBLG}

\appendix
\onecolumngrid
\vspace{1 cm}
\begin{center}
\large{\textbf{Supplementary material}}
\end{center}

\section{moir\'e Hamiltonian}
\label{app:MH}

\begin{center}
\begin{figure}
a) 
\includegraphics[scale=0.5]{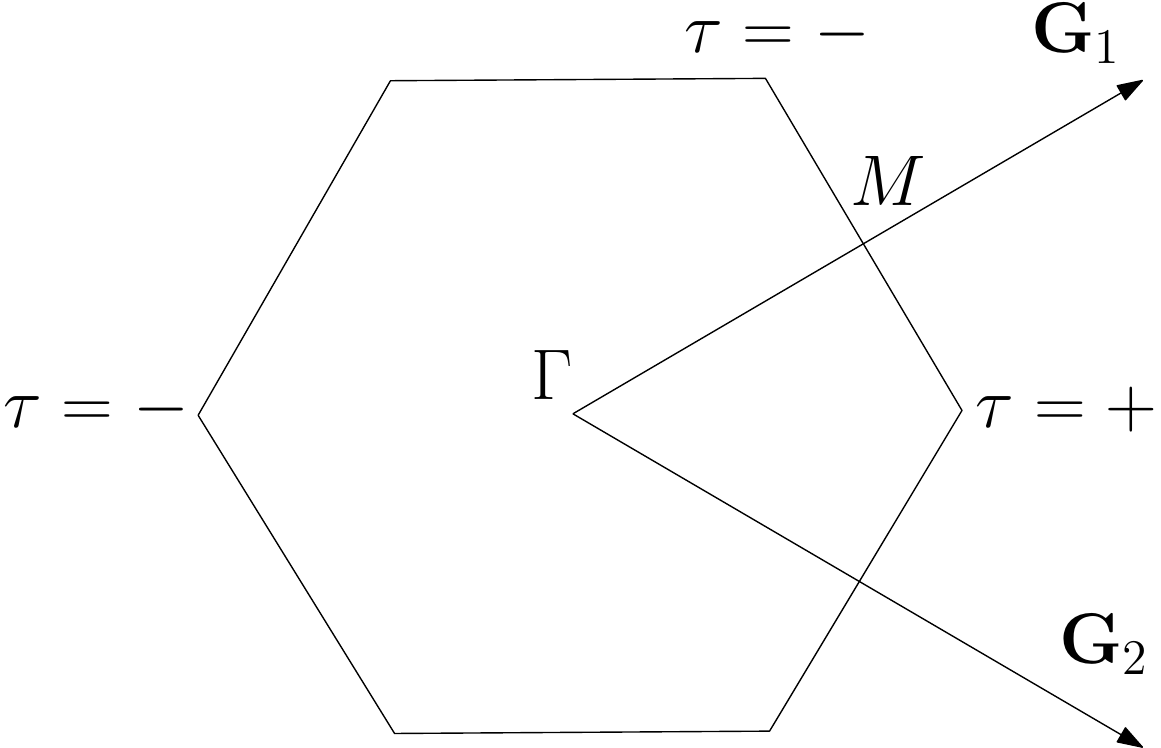} \hspace{1.5 cm} b)
\includegraphics[scale=0.6]{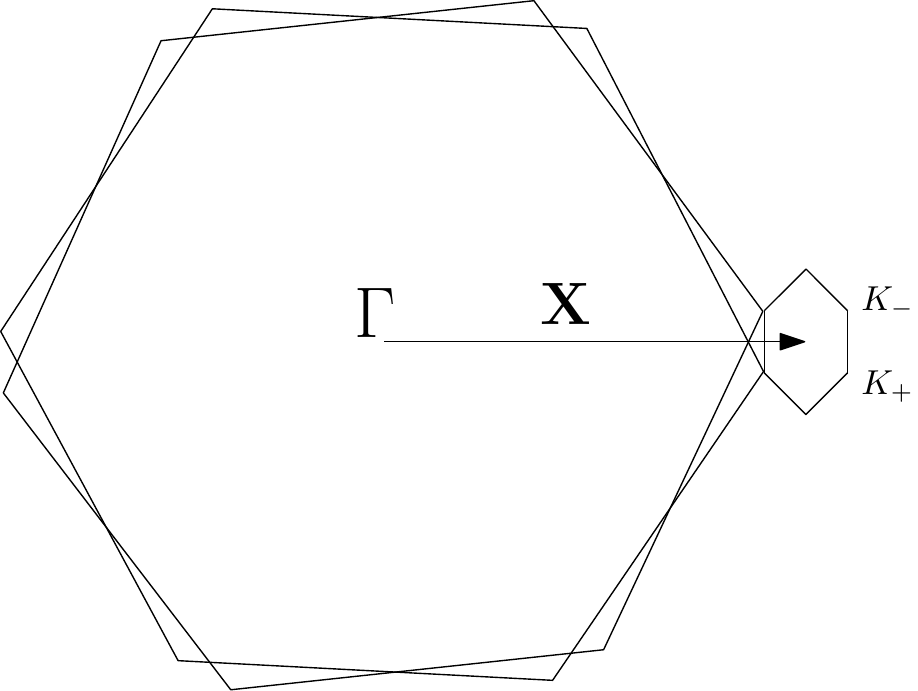}\caption{(a) The mono-layer graphene Brillouin zone with the two basis vectors $\textbf{G}_1$ and $\textbf{G}_2$ of the reciprocal lattice. We have indicated the high-symmetry $K$ points, where the Dirac cones are located, by the valley label $\tau = \pm$. (b) The mono-layer Brillouin zones of the top and bottom graphene layer with relative twist angle $\theta$. The vector $\textbf{X}$ points from the common $\Gamma$ point of the mono-layer Brillouin zones to the center of the mini-Brillouin zone at the $\tau = +$ valley. In presence of $C_6 T$ symmetry, there are Dirac points at the $K_+$ and $K_-$ points of the mini Brillouin zone (which is depicted by the small hexagon).}\label{fig:BZ}
\end{figure}
\end{center}

The spinless moir\'e Hamiltonian in valley $+$, i.e. around the $K_+$-points of the graphene Brillouin zone, is given by 

\begin{equation}
H(\textbf{k})=\sum_{\textbf{g}_1,\textbf{g}_2}\left(h^{tt}(R(\theta/2)(\textbf{k}+\textbf{X}+\textbf{g}_1))\delta_{\textbf{g}_1,\textbf{g}_2} + h^{bb}(R(-\theta/2)(\textbf{k}+\textbf{X}+\textbf{g}_1))\delta_{\textbf{g}_1,\textbf{g}_2} + \sum_{\tilde{\textbf{g}}}\left[ T_{\tilde{\textbf{g}}}^{tb} \delta_{\textbf{g}_1,\textbf{g}_2+\tilde{\textbf{g}}} + T_{\tilde{\textbf{g}}}^{bt} \delta_{\textbf{g}_1+\tilde{\textbf{g}},\textbf{g}_2} \right]\right) 
\end{equation}
Here, $\textbf{g}_1$ and $\textbf{g}_2$ lie on the moir\'e reciprocal lattice, $R(\pm\theta/2)$ is a rotation matrix over angle $\pm\theta/2$ with $\theta$ corresponding to the first magic angle $\theta\approx 1.05^\circ$ \cite{Bistritzer}. $h^{tt}(\textbf{k})= -t_0h(\textbf{k})+\Delta_t\sigma^z$ ($h^{bb}(\textbf{k})=-t_0 h(\textbf{k})+\Delta_b\sigma^z$) is the mono-layer graphene Hamiltonian of the top (bottom) layer with hopping strength $t_0=2.61$ eV and a sublattice splitting $\Delta_t\sigma^z$ ($\Delta_b\sigma^z$). $\textbf{X}$ is the position of the center of the mini-Brillouin zone at the mono-layer $K_+$-points as shown in Fig.\ref{fig:BZ}(b). The inter-layer coupling is given by the matrices \cite{Bistritzer}

\begin{eqnarray}
T_{\textbf{0}} &=&\left(\begin{matrix} w_0 & w_1 \\ w_1& w_0 \end{matrix}\right)\\
T_{\textbf{g}_1}&=&\left(\begin{matrix} w_0 & w_1\omega \\ w_1\omega^*& w_0 \end{matrix}\right) \\
T_{\textbf{g}_2}&=&\left(\begin{matrix} w_0 & w_1\omega^* \\ w_1\omega & w_0 \end{matrix}\right)\, ,
\end{eqnarray}
where $\omega=e^{i2\pi/3}$, $\textbf{g}_1 = (R(\theta/2)-R(-\theta/2))\textbf{G}_1$ and $\textbf{g}_2 = (R(\theta/2)-R(-\theta/2))\textbf{G}_2$, with $\G_1$ and $\G_2$ the graphene reciprocal lattice vectors shown in Fig. \ref{fig:BZ}. The AB inter-layer hopping strength is $w_1= 195$ meV. To phenomenologically incorporate corrugation of the bilayer system \cite{Wijk,Uchida,Lin,Lucignano} we use an AA-AB inter-layer hopping ratio $w_0/w_1=0.85$ \cite{Nam,Koshino2,Carr}. The moir\'e Hamiltonian in valley $-$ can be obtained by acting with time-reversal on the moir\'e Hamiltonian in valley $+$.

\section{Phonon Hamiltonian and electron-phonon coupling}
\label{app:elPh}

In this appendix, we review electron-phonon coupling in graphene, and phonon-mediated electron interactions in tBLG. The potential relevance of phonons for the superconducting domes and transport in magic-angle tBLG graphene was studied previously in Refs. \cite{WuMacDonald,LianBernevig,WuDasSarma,Choi,Polshyn}. Our approach to incorporate the effects of phonons is most closely related to that of Ref. \cite{WuMacDonald}, where mono-layer graphene phonons near both the $\Gamma$ and $K$ points were taken into account (these are the modes that couple most efficiently to the electrons \cite{FengWang,Sedeki,Piscanec}). In Refs. \cite{LianBernevig,WuDasSarma,KoshinoSon}, only long-wavelength acoustic phonons were considered. Here, we ignore these acoustic modes, as they do not give rise to inter-valley scattering for the electrons. The analysis below is solely based on the symmetry properties of graphene, and parallels the approach of Ref. \cite{BaskoAleiner}.

\subsection{Phonon Hamiltonian}

We define the Fourier transformed displacement operators $\hat{u}^i_{\q,\sigma}$ for the carbon atoms and the canonical conjugate operators $\hat{p}^i_{\q,\sigma}$ as

\begin{eqnarray}
\hat{u}^i_{\q, A} & = & \frac{1}{\sqrt{N}}\sideset{}{'}\sum_\q e^{i\q\cdot\R_A} \hat{u}^i_A(\R_A) \\
\hat{u}^i_{\q, B} & = & \frac{1}{\sqrt{N}}\sideset{}{'}\sum_\q e^{i\q\cdot(\R_A+\bdelta_1)} \hat{u}^i_B(\R_A+\bdelta_1) \\
\hat{p}^i_{\q, A} & = & \frac{1}{\sqrt{N}}\sideset{}{'}\sum_\q e^{-i\q\cdot\R_A} \hat{p}^i_A(\R_A) \\
\hat{p}^i_{\q, B} & = & \frac{1}{\sqrt{N}}\sideset{}{'}\sum_\q e^{-i\q\cdot(\R_A+\bdelta_1)} \hat{p}^i_B(\R_A+\bdelta_1)\, , 
\end{eqnarray}
where $i=x,y$, $\sigma$ denotes sublattice, $N$ is the number of unit cells, $\R_A$ denotes the positions of the $A$ sublattice sites, $\bdelta_1$ is one of the three vectors $\bdelta_l$ ($l=1,2,3$) pointing from the $A$ sublattice sites to the neighboring $B$ sublattice sites. Recall that we define primed sums to run over the graphene Brillouin zone. We only consider in-plane displacements, as the out-of-plane displacements couple only weakly to the electrons. Using the combined four-dimensional index $\nu=(i,\sigma)$, the phonon Hamiltonian can be written as

\begin{eqnarray}
H_{ph} & = & \frac{1}{2}\sideset{}{'}\sum_{\q}\left(\frac{1}{M}\sum_{\nu}\hat{p}_{\q,\nu}\hat{p}_{\q,\nu}^\dagger + 2\sum_{\nu,\nu'}\hat{u}_{\q,\nu}D(\q)_{\nu\nu'}\hat{u}^\dagger_{\q,\nu'} \right) \\
 & = & \frac{1}{2}\sideset{}{'}\sum_\q\left(\frac{1}{M}\sum_{\nu,\nu',j}\hat{p}_{\nu}\e^{j}_{\q,\nu}\e^{j*}_{\q,\nu'}\hat{p}^\dagger_{\nu'} + 2\sum_{\nu,\nu',j} \hat{u}_{\q,\nu}\e^{j*}_{\q,\nu}\lambda_{\q,j}\e^{j}_{\q,\nu'}\hat{u}_{\q,\nu'}^\dagger\right) \\
& = & \frac{1}{2}\sideset{}{'}\sum_\q\left( \frac{1}{M}\sum_{j} \hat{p}_{\q,j}\cdot \hat{p}_{\q,j}^\dagger+2\sum_j \hat{u}_{\q,j}\lambda_{\q,j} \hat{u}^\dagger_{\q,j} \right)\, ,
\end{eqnarray}
where $M$ is the carbon atom mass. Using $\omega_{\q,j}=\sqrt{2\lambda_{\q,j}/M}$, we define the phonon annihilation and creation operators as

\begin{eqnarray}
b_{\q,j} & = & \frac{i}{\sqrt{2M\hbar\omega_{\q,j}}}\hat{p}^\dagger_{\q,j}+\sqrt{\frac{\lambda_{\q,j}}{\hbar\omega_{\q,j}}}\hat{u}_{\q,j} \\
 b^\dagger_{\q,j} & = & \frac{-i}{\sqrt{2M\hbar\omega_{\q,j}}}\hat{p}_{\q,j}+\sqrt{\frac{\lambda_{\q,j}}{\hbar\omega_{\q,j}}}\hat{u}^\dagger_{\q,j}
\end{eqnarray}
In terms of the creation and annihilation operators, the phonon Hamiltonian becomes

\begin{equation}
H_{ph} = \sideset{}{'}\sum_\q\sum_j \hbar\omega_{\q,j}\left(b^\dagger_{\q,j}b_{\q,j}+\frac{1}{2}\right)
\end{equation}
Using the eigenvectors of the phonon Hamiltonian we can write the displacement operator in second quantization as

\begin{equation}
\hat{u}_\nu(\r)= \sideset{}{'}\sum_{\q,j} \sqrt{\frac{\hbar}{2NM\omega_{\q,j}}}(b_{\q j}+b^\dagger_{-\q j})\e^j_{\q \nu}e^{-i\q\cdot\r}\, ,
\end{equation}
For future convenience, we also introduce the notation

\begin{eqnarray}
b_{\q,j} & \equiv &  \langle e^j_{\q}|b_{\q}\rangle \\
& = & \sum_\nu \e^{j*}_{\q,\nu} b_{\q,\nu} \\
 & = & \sum_\nu \e^{j*}_{\q,\nu}\left(\frac{i}{\sqrt{2M\hbar\omega_{\q,j}}}\hat{p}^\dagger_{\q,\nu}+\sqrt{\frac{\lambda_{\q,j}}{\hbar\omega_{\q,j}}}\hat{u}_{\q,\nu} \right)
\end{eqnarray}

\subsection{Electron-phonon coupling in graphene}

In a tight-binding approximation, the only coupling between electrons and lattice vibrations occurs via the associated spatial modulation of the tight binding parameters. In the case of graphene we write the tight-binding Hamiltonian coupled to small lattice vibrations as \cite{Viet,Suzuura,Ishikawa,Sasaki,Mahan}

\begin{eqnarray}\label{eq:tightb}
H & = & -t_0\sum_{\R_A}\sum_{l=1}^3\psi^\dagger_{\R_A}\psi_{\R_A+\bdelta_l} - \frac{\partial t_0}{\partial a_{CC}}\sum_{\R_A}\sum_{l=1}^3\left(|\bdelta_l + \u_A(\R_A) - \u_B(\R_A+\bdelta_l)| - a_{CC} \right)\psi^\dagger_{\R_A}\psi_{\R_A+\bdelta_l}  + h.c. \nonumber \\
& \approx & -t_0\sum_{\R_A}\sum_{l=1}^3\psi^\dagger_{\R_A}\psi_{\R_A+\bdelta_l} - \frac{1}{a_{CC}}\frac{\partial t_0}{\partial a_{CC}}\sum_{\R_A}\sum_{l=1}^3\bdelta_l\cdot(\u_A(\R_A) - \u_B(\R_A+\bdelta_l)) \psi^\dagger_{\R_A}\psi_{\R_A+\bdelta_l}  + h.c.
\end{eqnarray}
where $t_0$ is the graphene hopping strength, $a_{CC}=|\bdelta_l|$ the distance between two carbon atoms. Going to momentum space, the electron-phonon coupling Hamiltonian becomes

\begin{eqnarray}
H_{e-ph} & = & -\frac{1}{a_{CC}}\frac{\partial t_0}{\partial a_{CC}}\sideset{}{'}\sum_{\k,\q}\sum_{l=1}^3\bdelta_l \cdot\left(\u_A(\q)-\u_B(\q)e^{-i\q\cdot \bdelta_l} \right)e^{-i\k\cdot \bdelta_l}\psi^\dagger_{\k+\q,A}\psi_{\k,B} + h.c. \\ 
 & = & -\frac{1}{a_{CC}}\frac{\partial t_0}{\partial a_{CC}}\sideset{}{'}\sum_{\k,\q}\sum_j\sum_{l=1}^3\sqrt{\frac{\hbar}{2NM\omega_{\q,j}}}\bdelta_l\cdot\left(\e^j_{\q,A}-\e^j_{\q,B}e^{-i\q\cdot \bdelta_l} \right)e^{-i\k\cdot \bdelta_l}\psi^\dagger_{\k+\q,A}\psi_{\k,B}(b_{\q j}+b^\dagger_{-\q j}) \nonumber
\end{eqnarray}
By defining the vectors

\begin{equation}
|V_{\q,\k}\rangle = \sum_{l=1}^3\left(\bdelta_l e^{i\k\cdot\bdelta_l},-\bdelta_l e^{i(\k+\q)\cdot\bdelta_l} \right)\;,\;\;\;\;
|e^j_\q\rangle =  \left( \e^j_{\q,A},\e^j_{\q,B}\right)
\end{equation}
we can write the electron-phonon coupling Hamiltonian as

\begin{equation}\label{ham0}
H_{e-ph} = -\tilde{g}\sum_j \sideset{}{'}\sum_{\q,\k}\omega_{\q,j}^{-1/2}\langle V_{\q,\k}|e^j_\q\rangle\, \psi^\dagger_{\k+\q,A}\psi_{\k,B}(b_{\q j}+b^\dagger_{-\q j}) + h.c.
\end{equation}
where $\tilde{g}=\frac{1}{a_{CC}}\sqrt{\frac{\hbar}{2NM}}\frac{\partial t_0}{\partial a_{CC}}$. Let us now examine how the symmetries of graphene are realized in this Hamiltonian. We first consider the three-fold rotation symmetry group $C_{3v}$ and define the rotation matrix  $R_3$

\begin{equation}
R_3=\left(\begin{matrix}\cos(2\pi/3) & \sin(2\pi/3) \\ -\sin(2\pi/3) & \cos(2\pi/3) \end{matrix}\right)=\left(\begin{matrix}-1/2 & \sqrt{3}/2 \\ -\sqrt{3}/2 & -1/2\end{matrix}\right)\, .
\end{equation}
$C_{3v}$ symmetry of the phonon Hamiltonian implies that

\begin{equation}
D(R_3\q) = R^\dagger D(\q)R\, ,\;\;\; \text{ with } R=\left( \begin{matrix}R_3 & \\ & R_3 \end{matrix}\right)\, ,
\end{equation}
from which it follows that $\omega_{\q,j}=\omega_{R_3\q,j}$ and $R|e^j_{\q}\rangle = e^{i\alpha_\q}|e^j_{R_3\q}\rangle$. The $C_{3v}$ symmetry of the electron-phonon Hamiltonian implies that

\begin{equation}
\omega^{-1/2}_{R_3\q, j} \langle V_{R_3\q,R_3\k}|e^j_{R_3\q}\rangle \langle e^j_{R_3\q}|R|b_\q\rangle = \omega^{-1/2}_{\q, j} \langle V_{\q,\k}|e^j_{\q}\rangle \langle e^j_{\q}|b_\q\rangle\, ,
\end{equation}
Because $\omega_{R_3\q,j}=\omega_{\q,j}$, we can see that this is true by doing following steps

\begin{eqnarray}
\langle V_{R_3\q,R_3\k}|e^j_{R_3\q}\rangle \langle e^j_{R_3\q}|R|b_\q\rangle & = & \langle V_{R_3\q,R_3\k}|RR^\dagger|e^j_{R_3\q}\rangle \langle e^j_{R_3\q}|R|b_\q\rangle \\
 & = & e^{-i\alpha_\k}\langle V_{R_3\q,R_3\k}|R|e^j_{R_3\q}\rangle \langle e^j_{\q}|b_\q\rangle e^{i\alpha_\k}\\
 & = & \langle V_{\q,\k}|e^j_{\q}\rangle \langle e^j_{\q}|b_\q\rangle
\end{eqnarray}
The $C_{2v}$ symmetry can be derived in a similar way, with the main difference that $C_{2v}$ interchanges the $A$ and $B$ sublattices. So $C_{2v}$ symmetry implies that

\begin{equation}\label{C2}
\omega^{-1/2}_{-\q,j}\langle V_{-\q,-\k}|e^j_{-\q}\rangle \langle e^j_{-\q}|\tilde{R}|b_{\q}\rangle = \omega^{-1/2}_{-\q,j}\langle V_{-\q,\k+\q}|e^j_{-\q}\rangle^* \langle e^j_{\q}|b_{\q}\rangle\, ,\;\;\;\text{ with } \tilde{R}=\left(\begin{matrix} & - \mathds{1} \\ -\mathds{1} &  \end{matrix}\right)
\end{equation}
Equality \eqref{C2} follows from the definition of $|V_{\k,\q}\rangle$, the $C_{2v}$ rotation symmetry of the phonon Hamiltonian which implies that $\tilde{R}|e^j_\q\rangle = e^{i\beta_\q}|e^j_{-\q}\rangle$, and $|e^j_{\q}\rangle = |e^j_{-\q}\rangle^*$, which follows from hermiticity of the displacement operator. Time reversal symmetry of the electron-phonon Hamiltonian in Eq. \eqref{ham0} is more straightforward to see, as this simply follows from the properties $|V_{-\q,-\k}\rangle^*=|V_{\q,\k}\rangle$ and $|e^{j}_{-\q}\rangle^*=|e^{j}_{\q}\rangle$.

We now focus on the coupling between lattice-scale phonons and low-energy electrons at the Dirac cones. So in the above electron-phonon Hamiltonian we fix both $\k$ and $\q$ to either $\K$ or $-\K$, where $\K=\left(\frac{4\pi}{3a},0 \right)$ and $a=\sqrt{3}a_{CC}$ is the graphene lattice constant. Specifically, the terms we are interested in are

\begin{eqnarray} 
H_{e-ph} & \approx &  -\tilde{g} \sum_j \omega_{\K,j}^{-1/2} \langle V_\K|e^j_{\K}\rangle \psi^\dagger_{-\K,A}\psi_{\K,B}(b_{\K j}+b^\dagger_{-\K j}) \\
 & &   -\tilde{g} \sum_j \omega_{\K,j}^{-1/2} \langle V_\K|e^j_{\K}\rangle^* \psi^\dagger_{\K,A}\psi_{-\K,B}(b_{-\K j}+b^\dagger_{\K j}) + h.c. \, ,\nonumber
\end{eqnarray}
where $|V_\K\rangle = |V_{\K,\K}\rangle$. Let us now choose a basis in which the $\bdelta_l$ take the form 

\begin{equation}
\bdelta_1  =  a_{CC}(0,1)\,,\;\;\; \bdelta_2 = a_{CC}\left(\frac{\sqrt{3}}{2},-\frac{1}{2} \right) = R_3\bdelta_1\,,\;\;\;\;\bdelta_3 = a_{CC}\left(-\frac{\sqrt{3}}{2},-\frac{1}{2} \right) = R_3\bdelta_2
\end{equation}
from which we see that $e^{i\K\cdot\bdelta_1}=1$, $e^{i\K\cdot\bdelta_2}=e^{2\pi i/3}\equiv \omega$ and $e^{i\K\cdot\bdelta_3}= \omega^2=\omega^{-1}$. The phonon Hamiltonian at the $K_+$ point satisfies

\begin{equation}
R^\dagger D(\K) R = D(R_3\K)=D(\K-\G_2) = \left(\begin{matrix}\mathds{1} & \\ & e^{-i\G_2\cdot \bdelta_1}\mathds{1}  \end{matrix}\right)D(\K) \left(\begin{matrix}\mathds{1} & \\ & e^{i\G_2\cdot \bdelta_1}\mathds{1}  \end{matrix}\right)\, ,
\end{equation}
where we have used that $R_3\K = \K - \G_2$, with $\G_2 =\frac{4\pi}{\sqrt{3}a}\left(\frac{\sqrt{3}}{2},\frac{1}{2} \right)$ a reciprocal lattice vector. The last equality follows from $b_{\q+\G,A,x^i}=b_{\q,A,x^i}$ and $b_{\q+\G,B,x^i}=e^{i\G\cdot\bdelta_1}b_{\q,B,x^i}$ for any reciprocal lattice vector $\G$. Using that $e^{i\G_2\cdot \bdelta_1} = \omega$, we see that the matrix $R_- = R_3\oplus \omega^{-1}R_3$ commutes with $D(\K)$. This means that the eigenvectors $\e^j_{\K}$ are also eigenvectors of $R_-$, which has two non-degenerate eigenvalues $1$ and $\omega^{-1}$, and one two-fold degenerate eigenvalue $\omega$. The vector $|V_\K\rangle$ can be written as $|V_\K\rangle = |V^A_\K\rangle + |V^B_\K\rangle$, where

\begin{equation}
|V^A_\K\rangle = \sum_{l=1}^3 \left( \bdelta_l e^{i\K\cdot \bdelta_l},0\right) \;,\;\;\;\;|V^B_\K\rangle = \sum_{l=1}^3 \left( 0, -\bdelta_l e^{i2\K\cdot \bdelta_l}\right)
\end{equation}
These vectors have the property $R_-|V^A_\K\rangle = \omega^{-1}|V^A_\K\rangle$ and $R_-|V^B_\K\rangle = |V^B_\K\rangle$. This means that only two of the four inner products $\langle V_\K|e^j_\K\rangle$ are non-zero. The eigenvectors $|e^j_\K\rangle$ which can couple to the electrons are those which have eigenvalue $1$ and $\omega^{-1}$ under $R_-$. We can thus express $|V_{\K}\rangle$ in terms of the eigenvectors $|e^j_{\K}\rangle$ as follows:

\begin{equation}
\frac{1}{\sqrt{6}a_{CC}}|V_{\K}\rangle = \frac{1}{\sqrt{2}}(e^{i\theta^1_\K}|e^1_{\K}\rangle +e^{i\theta^2_\K}|e^2_{\K}\rangle)\, ,
\end{equation}
This allows us to write the electron-phonon Hamiltonian as

\begin{equation}
H_{e-ph} = -\tilde{g}\sqrt{3}a_{CC} \sum_{j=1}^2  \frac{e^{i\theta^j_\K}}{\sqrt{\omega_{\K j}}} \psi^\dagger_{-\K,A}\psi_{\K,B}(b_{\K,j}+b^\dagger_{-\K j}) + \frac{e^{-i\theta^j_\K}}{\sqrt{\omega_{\K j}}} \psi^\dagger_{\K,A}\psi_{-\K,B}(b_{-\K,j}+b^\dagger_{\K j}) + h.c.
\end{equation}
From $C_{2v}$ symmetry we know that $e^{i\theta^j_\K}=e^{i\theta^j_{-\K}}=e^{-i\theta^j_\K}$, which implies that $e^{i\theta^j_\K}$ is real and can be absorbed in $b_{\K,j}$ and $b^\dagger_{-\K,j}$. So the final form for the electron-phonon coupling between lattice-scale phonons and low-energy electrons at the $K$ points is simply

\begin{equation}\label{finalepc}
H_{e-ph} = -g \sum_{j=1}^2 \frac{1}{\sqrt{\omega_{\K,j}}}\left(\psi^\dagger_{-\K,A}\psi_{\K,B}(b_{\K,j}+b^\dagger_{-\K j}) +  \psi^\dagger_{\K,A}\psi_{-\K,B}(b_{-\K,j}+b^\dagger_{\K j}) \right) + h.c.\, ,
\end{equation}
where $g = \sqrt{\frac{3\hbar}{2NM}}\frac{\partial t_0}{\partial a_{CC}}$. Because the graphene phonon bands have little dispersion around the $K$-points \cite{Gruneis,Suzuura,Yan}, we will now simply ignore any momentum dependence and simply assume that \eqref{finalepc} holds for electrons close to the $K$-points. We will also take $\omega_{\K,1}=\omega_{\K,2}=\omega_0$.

\subsection{Phonon mediated electron interactions}

The Hamiltonian describing the combined electron-phonon system, projected into the flat bands, takes the form

\begin{equation}
H = H_e + H_{ph} + H_{e-ph}\, ,
\end{equation}
with $H_e = \sum_{\k,\tau,s} \varepsilon_{\k,\tau}c^\dagger_{\k,\tau,s}c_{\k,\tau,s}$. For the phonon Hamiltonian we take just two copies of the graphene phonon Hamiltonian:

\begin{eqnarray}
H_{ph} & = & \sum_{\q,\g}\sum_{l,j} \hbar\omega_{\q+\g,l,j}\left(b^\dagger_{\q+\g,l,j}b_{\q+\g,l,j}+\frac{1}{2}\right)\,,
\end{eqnarray}
where $\q$ is defined to lie in the mini-Brillouin zone. We don't consider out-of-plane phonon modes as these couple only to the inter-layer tunneling, which is much smaller than the intra-layer hopping. Correspondingly, the electron-phonon Hamiltonian is just two copies of Eq. \eqref{finalepc}. If we project this into the flat bands, we get

\begin{eqnarray}
H_{e-ph} & = & -\frac{g}{\sqrt{\omega_0}}\sum_{\tau,l,j,\g}\sum_{\k,\q,s} \langle u_{-\tau}(\k+\q)|\sigma^x P_lS_\g|u_\tau(\k)\rangle  c^\dagger_{\k+\q,-\tau,s}c_{\k,\tau,s}\left(b_{\q+\g+2\tau\X,l,j}+b^\dagger_{-\q-\g-2\tau\X,l,j} \right)\\
 & \equiv & -\frac{g}{\sqrt{\omega_0}}\sum_{l,j,\g}\sum_{\k,\q,\tau,s} f_{l,\g}^{\tau}(\q,\k) c^\dagger_{\k+\q,-\tau,s}c_{\k,\tau,s}\left(b_{\q+\g+2\tau\X,l,j}+b^\dagger_{-\q-\g-2\tau\X,l,j} \right)\, ,
\end{eqnarray}
Using a Schrieffer-Wolff transformation we obtain following phonon-mediated electron interaction Hamiltonian

\begin{eqnarray}\label{twistedepc}
H_{PH} & = & \frac{2g^2}{\omega_0}\sum_{\k,\k',\q}\sum_{\tau,s,s'}\sum_{l}\hbar\omega_{0}\frac{f^{\tau}_{l,\g}(\k,\q)f^{-\tau}_{l,-\g}(\k',-\q)}{(\varepsilon_{\k+\q,-\tau}-\varepsilon_{\k,\tau})^2-(\hbar\omega_{0})^2}c^\dagger_{\k+\q,-\tau,s}c_{\k,\tau,s} c^\dagger_{\k'-\q,\tau,s'}c_{\k',-\tau,s'} \nonumber \\
 & \approx & -\frac{2g^2\hbar}{(\hbar\omega_0)^2} \sum_{\k,\k',\q}\sum_{s,s'}\sum_{\tau}\left(\sum_{\l,\g} f^{\tau}_{l,\g}(\k,\q)f^{-\tau}_{l,-\g}(\k',-\q) \right)\,c^\dagger_{\k+\q,-\tau,s}c_{\k,\tau,s} c^\dagger_{\k'-\q,\tau,s'}c_{\k',-\tau,s'} \,, \nonumber
\end{eqnarray}
where we have again ignored the phonon dispersion, and also the flat band dispersion. The interaction strength $g_{ph}$ used in the main text is

\begin{equation}
g_{ph} = \frac{3\hbar^2}{2M}\frac{\beta^2}{(\hbar\omega_0)^2}\left(\frac{t_0}{a_{CC}} \right)^2\, ,
\end{equation} 
where $\beta = \partial \ln t_0/\partial \ln a_{CC}$. The numerical value $g_{ph}\approx 630$ meV can be obtained by using $\hbar\omega_0 = 0.16$ eV, $t_0=2.61$ eV, $a_{CC}=0.25/\sqrt{3}$ nm and $\beta=3$ \cite{Ribeiro,WuMacDonald}.

\section{Spin stiffness in a spin polarized flat Chern band}
\label{app:rhos}

In this section we derive an expression for the spin stiffness associated with a spin polarized flat Chern band. The spin stiffness $\rho_s$ appears in a long-wavelength description as the coefficient of the gradient term in the effective action describing spin fluctuations:

\begin{equation}
\frac{\rho_s}{2} \int\mathrm{d}\r\, (\nabla \n)^2
\end{equation}
To derive $\rho_s$ within mean-field theory, we generalize the calculation of Ref. \cite{Moon} for a spin-polarized lowest Landau level to a Chern insulator. We assume that in the ground state the spins are polarized in the $z$-direction. We create a non-homogeneous spin texture by acting with $e^{i\hat{O}}$ on the uniformly polarized ground state wave function. The operator $e^{i\hat{O}}$ is defined as

\begin{equation}
e^{i\hat{O}} = e^{i \sum_\r \bOmega(\r)\cdot \S(\r)} = e^{i\sum_\q \bOmega(\q)\cdot \S(-\q)} \, ,
\end{equation}
where $\S(\r)$ is the spin operator at site $\r$. We will assume that the resulting spin texture consists only of small fluctuations around the $z$-direction, such that $\bOmega(\r) \approx \hat{z}\times \n(\r)$, and is slowly varying in space. If we project $e^{i\hat{O}}$ in a Chern band with band label $\mu$, the resulting operator $e^{i\hat{O}_\mu}=e^{i\sum_\q \bOmega(\q)\cdot \S_{\mu}(-\q)}$ is defined using the projected spin operator

\begin{equation}\label{spinprojected}
\S_\mu(-\q) = \frac{1}{\sqrt{N}}\sum_\k \langle u_\mu(\k-\q)|u_\mu(\k)\rangle c^\dagger_{\k-\q,\mu}\frac{\s}{2}c_{\k,\mu} \equiv \frac{1}{\sqrt{N}}\sum_\k \lambda_{\mu}(-\q,\k) c^\dagger_{\k-\q,\mu}\frac{\s}{2}c_{\k,\mu}\, ,
\end{equation}
where the operator $c^\dagger_{\k,\mu}$ creates an electron with crystal momentum $\k$ in band $\mu$, $N$ is the number of unit cells, $\s=(s^x,s^y,s^z)$ are the Pauli spin operators, and $|u_\mu(\k)\rangle$ are the periodic Bloch states. From now on, we will drop the band index $\mu$. This should not cause any confusion, as we are always considering the same single band.

We are interested in the energy increase associated with the spin texture in the small $|\q|$ limit, which we get from

\begin{eqnarray}
\delta E & = & \langle e^{i\hat{O}}H e^{-i\hat{O}}\rangle - \langle H \rangle \\ 
 & = & i \langle [\hat{O},H]\rangle -\frac{1}{2}\langle [\hat{O},[\hat{O},H]]\rangle +\cdots
\end{eqnarray}
For the Hamiltonian we use a general density-density interaction $\sum_\k \tilde{V}(\k):\rho(\k)\rho(-\k):$\,, projected into the flat Chern band. So the commutator we need to calculate is

\begin{equation}
[\hat{O},H] = \sum_{\k,\q} \sum_i \Omega^i(\q)\tilde{V}(\k)[S^i(-\q),\rho(\k)\rho(-\k)]
\end{equation}
We can easily evaluate this by applying the identity

\begin{equation}
[S^i(-\q),\rho(\k)\rho(-\k)] = [S^i(-\q),\rho(\k)]\rho(-\k) + \rho(\k)[S^i(-\q),\rho(-\k)]
\end{equation}
Using the explicit expression $\rho(\k)=\frac{1}{\sqrt{N}}\sum_{\k'}\lambda(\k,\k')c^\dagger_{\k'+\k}c_{\k'}$ for the projected density operator, and Eq. \eqref{spinprojected}, we find

\begin{eqnarray}
[S^i(-\q),\rho(\k)] & = &  \frac{1}{N}\sum_{\k'}\left(\lambda(\k,\k')\lambda(-\q,\k+\k')-\lambda(\k,\k'-\q)\lambda(-\q,\k')\right)c^\dagger_{\k'+\k-\q}\frac{s^i}{2}c_{\k'} \\
 & \equiv & \frac{1}{N}\sum_{\k'}\Lambda_{\k',\k,-\q}c^\dagger_{\k'+\k-\q}\frac{s^i}{2}c_{\k'} 
\end{eqnarray}
and thus

\begin{equation}
[\hat{O},H] = \frac{1}{N}\sum_{i,\k,\q}\Omega^i(\q)\tilde{V}(\k)\sum_{\k'}\left( \Lambda_{\k',\k,-\q}c^\dagger_{\k'+\k-\q}\frac{\sigma^i}{2}c_{\k'}\rho(-\k) + \Lambda_{\k',-\k,-\q}\rho(\k) c^\dagger_{\k'-\k-\q}\frac{\sigma^i}{2}c_{\k'}\right)
\end{equation}
The expectation value of this commutator with respect to the homogeneously $z$-polarized Slater determinant vanishes because $\Omega^z=0$. 

The double commutator determining the energy change in second order becomes

\begin{eqnarray}
[\hat{O},[\hat{O},H]] & = & \frac{1}{N}\sum_{i,j}\sum_{\k,\q_1,\q_2}\Omega^i(\q_1)\Omega^j(\q_2)\tilde{V}(\k)\times \nonumber \\
 & & \sum_{\k'} \bigg( \Lambda_{\k',\k,-\q_1}[S^j(-\q_2),c^\dagger_{\k'+\k-\q_1}\frac{s^i}{2}c_{\k'}\rho(-\k)] +\Lambda_{\k',-\k,-\q_1}[S^j(-\q_2),\rho(\k)c^\dagger_{\k'-\k-\q_1}\frac{s^i}{2}c_{\k'}]\bigg)
\end{eqnarray}
Evaluating the expectation value of this double commutator is tedious, but straightforward. We find

\begin{eqnarray}
\langle [\hat{O},[\hat{O},H]]\rangle & = &\frac{1}{N^{2}}\sum_{i,\k,\q}\Omega^i(\q)\Omega^i(-\q)\tilde{V}(\k) \nonumber \\
 & & \times \sum_{\k'}\Lambda_{\k',\k,-\q}\left(\lambda(\q,\k'+\k-\q)\lambda(-\k,\k'+\k) - \lambda(\q,\k'-\q)\lambda(-\k,\k'+\k-\q)  \right)
\end{eqnarray}
To simplify the product of form factors $\lambda$, we work up to second order in $\q$, because by assumption $\bOmega(\q)$ is a fast decaying function. The interaction $V(\k)$ is in general not decaying fast enough to justify working up to second order in $\k$. However, the expectation value of the double commutator contains factors of the form $\lambda(\k,\k')=\langle u(\k+\k')|u(\k')\rangle$, which are expected to decay very fast in $|\k|$. So this decay does allow us to work up to second order in $\k$, but we need to explicitly keep the function $f(\k,\k') = |\lambda(\k,\k')|$. We expect the decay of the form factors not to vary too much over the Brillouin zone, so we will use the function $f(\k) = |\lambda(\k,\k_0)|$ for a fixed representative $\k_0$ in the Brillouin zone to enforce the fast decay in $|\k|$ (for example, Ref. \cite{YaHuiZhang1} chose $\k_0 = 0$). The Taylor expanded expressions for the form factors contain a term proportional to the Berry connection, which provides the connection between a Landau level and a Chern band, as noted in Ref. \cite{Parameswaran}. After a few straightforward manipulations, we find for the energy difference

\begin{eqnarray}
\delta E & = &  \frac{1}{8N^{2}}\sum_{i,\k,\q}\Omega^i(\q)\Omega^i(-\q)\tilde{V}(\k) (\q\wedge\k)^2 \sum_{\k'}\mathcal{F}(\k')^2 f^2(\k) \\
 & = & \frac{1}{16}\left(\frac{1}{N}\sum_{\k'}\mathcal{F}(\k')^2\right) \left(\frac{1}{N}\sum_\k \tilde{V}(\k)f^2(\k)|\k|^2 \right)\sum_{i,j,\q}(iq^j\Omega^i(\q))(-iq^j\Omega^i(-\q)) \\
 & = & \frac{1}{16}\left(\frac{1}{N}\sum_{\k'}\mathcal{F}(\k')^2\right) \left(\frac{1}{N}\sum_\k \tilde{V}(\k)f^2(\k)|\k|^2 \right)\sum_{i,\r} (\nabla\Omega^i(\r))\cdot(\nabla\Omega^i(\r))\\
 & \rightarrow & \frac{\rho_s}{2}\int\mathrm{d}^2\r\,(\nabla \n)^2\, ,
\end{eqnarray}
where in the second line we have used $(\q\wedge\k)^2=|\q|^2|\k|^2\sin^2\alpha$, where $\alpha$ is the angle between $\q$ and $\k$. Because $\tilde{V}(\k)$ and to a good approximation also $f(\k)$ are isotropic, we can replace $\sin^2\alpha$ by its average value $1/2$. So we arrive at the following Hartree-Fock expression for the spin stiffness

\begin{equation}
\rho_s = \frac{1}{8A}\left(\frac{1}{N}\sum_{\k'}\mathcal{F}(\k')^2\right) \left(\frac{1}{N}\sum_\k \tilde{V}(\k)f^2(\k)|\k|^2 \right)\, ,
\end{equation}
where $A$ is the area of the unit cell. In the continuum limit, the factor $A^{-1}$ is interpreted as the charge density \cite{Girvin}. 

\section{Skyrmion energetics}
\subsection{Single skyrmions}
\label{app:EnSky}
In this section, we present an explicit evaluation of the energy of a skyrmion in a single-valley, using the two-component non-linear $\sigma$ model discussed in Eq.~(\ref{eq:Leff}), which we recall below for completeness. We assume that while a skyrmion forms in a single valley, the spins in the other valley remain in their equilibrium configuration. We first look at the ferromagnet.
\beq
\mathcal{L} =  \sum_{\tau=\pm} \left[ n S \bigg( \bm{A}[\n_\tau] \cdot \partial_t \n_\tau(\r)  + g_s \mu_B \B \cdot \n_\tau(\r) \bigg) - \frac{\rho_s}{2} (\nabla \n_\tau (\r))^2 \right] - \frac{n S^2 \bro}{2} [(\n_+(\r) - \n_-(\r)]^2- \frac{1}{2} \int d\rp \, V(\r - \rp) \rho(\r) \rho(\rp) \nn
 \label{eq:LeffFMApp}
 \eeq
We henceforth set $S = 1/2$ for the electron spin. We consider a single isolated skyrmion in valley $+$ (say), completely characterized by a complex function $W(z)$ (see Eq.~(\ref{eq:Wans})) As shown by Belavin and Polyakov, any analytic complex function $W(z)$ with a single pole minimizes the elastic energy $E^{\text{el}}$ to be $4\pi\rho_s$ \cite{BP75}, and the size of a charged skyrmion in a Chern band is therefore determined by the competition between the effective Zeeman and Coulomb energies \cite{Sondhi}. A skyrmion of linear size $R$ can be described by $W(z) = R/z$, or more explicitly by
 \beq
 \n_+(\r) = \left( \frac{2 x R}{r^2 + R^2}, \frac{2 y R}{r^2 + R^2}, \frac{r^2 - R^2}{r^2 + R^2} \right), \text{ and } \n_-(\r) = (0,0,1)
 \label{eq:SkyAns}
 \eeq
 We want to optimize the size $R$ as a function of the ratio of effective Zeeman energy $\Delta$ (which is a combination of the external magnetic field $\bm{B} = B_\perp \hz$ and the internal exchange field from the other valley $\bro \; \n_-(\r) = \bro \; \hz$) to the Coulomb energy, i.e, $\tilde{g}$ which we define below. \beq
\tilde{g} \equiv \frac{\Delta}{E_C} = \frac{g_s \mu_B \tilde{B}}{\frac{e^2}{4\pi \epsilon a_M}}, \text{ where } \tilde{B} =  B_\perp + \frac{\bro}{2 g_s \mu_B} 
 \eeq
 If we naively use the effective Hamiltonian from Eq.~(\ref{eq:LeffFMApp}) to compute the energy, the Zeeman term will diverge as a very large number of spins are flipped in our ansatz in Eq.~(\ref{eq:SkyAns}). There is a natural cutoff set by the correlation length of spin fluctuations, as the Goldstone mode in a single valley gets gapped in presence of the effective magnetic field $\tilde{B}$. In particular, we can use the equation of motion derived from Eq.~(\ref{eq:LeffFMApp}) to get the dispersion of a neutral spin-wave.
 \beq
 \frac{\partial \n_+}{\partial t} &=& \left( \frac{2 \rho_s}{n} \nabla^2 \n_+ +  g_s \mu_B \tilde{B} \hat{z} \right) \times \n_+ \implies i  \frac{\partial \psi_+}{\partial t} = \left(\frac{2\rho_s}{n} \nabla^2 - g_s \mu_B \tilde{B} \right) \psi_+ \text{ where } \psi_+ = n_{+,x} + i n_{+,y} \nn
 \implies \omega_\k &=& \frac{2\rho_s}{n} k^2 + g_s \mu_B \tilde{B} \equiv \frac{2\rho_s}{n}(k^2 + \xi_s^{-2})
 \eeq
 This implies that the spin-correlations fall off exponentially beyond a length-scale set by $\frac{\xi_s}{a_M} \equiv \left( \frac{\sqrt{3} \rho_s}{g_s \mu_B \tilde{B}} \right)^{1/2} \approx \left( \frac{E_C}{\Delta} \right)^{1/2}$, where $n = \frac{2}{\sqrt{3}a_M^2}$ is the density of electrons per band, and we have used that the spin-stiffness $\rho_s$ is set by the Coulomb energy scale $E_C = \frac{e^2}{4\pi \epsilon a_M}$. Note that we treat $\epsilon$ as a phenomenological dielectric constant that also takes into account the renormalization of the bare Coulomb energy due to projection to the relevant flat bands. Therefore, we can write down the total excitation energy of the skyrmion as the sum of the elastic contribution $E_{el}$, the effective Zeeman contribution $E_Z$ and the Coulomb contribution. 
 \beq
 E_{sk} = 4 \pi \rho_s + \frac{g_s \mu_B \tilde{B}}{\sqrt{3} a_M^2} \int_{0}^{\kappa \xi_s} d^2 r \; [1 - n_z(\r)] + \frac{1}{2} \int \frac{d^2q}{(2 \pi)^2} V(\q) \rho_{\q} \rho_{-\q}
 \label{eq:Es}
 \eeq
 The first term, which is the elastic contribution, is independent of the size of the skyrmion \cite{BP75}. The effective Zeeman energy, with a cutoff $\kappa \xi_s$ for the domain of integration is given by (the additional scale factor of $\kappa$ is added for later analytical convenience)
\beq
E_Z = \frac{g_s \mu_B \tilde{B}}{\sqrt{3} a_M^2} \int_{0}^{\kappa \xi_s} d^2 r \; [1 - n_z(\r)]  = \frac{2 \pi g_s \mu_B \tilde{B} R^2}{\sqrt{3} a_M^2} \ln \left( \frac{R^2 + (\kappa \xi_s)^2}{R^2} \right)
\eeq
We first discuss the case of unscreened Coulomb interaction $V(\r) = \frac{1}{4 \pi \epsilon r}$, as would be expected for a dilute gas of skyrmions in the absence of gate-screening. Therefore, we take $V(\q) = \int d^2r \, V(\r) e^{i \q \cdot \r} = \frac{1}{2 \epsilon q}$ and compute the Coulomb energy
\beq
\rho(\r) = -\frac{1}{8\pi} \epsilon_{\mu \nu} \n \cdot (\partial_\mu \n \times \partial_\nu \n) = - \frac{R^2}{\pi (r^2 + R^2)^2} \implies \rho_\q = \int d^2r \; \rho(\r) e^{i \q \cdot \r} = - q R \, K_1(q R) \nn
\implies \frac{1}{2} \int \frac{d^2q}{(2 \pi)^2} V(\q) \rho_{\q} \rho_{-\q} = \frac{e^2}{8 \pi \epsilon  R} \int_{0}^\infty dt \; \left[ t K_1(t) \right]^2 =  \frac{3 \pi e^2}{2^{8}  \epsilon R}
\eeq
Now, we parametrize the size of the skyrmion by $R = \kappa a_M$ (roughly speaking, $\kappa^2$ counts the number of flipped spins), and minimize the skyrmion energy $E_{sk}$ in Eq.~(\ref{eq:Es}) as a function by $\kappa$.
\beq
E_{sk}(\kappa) = 4 \pi \rho_s + \frac{2 \pi}{\sqrt{3}} \Delta \kappa^2 \ln \left( 1+ \frac{E_C}{\Delta} \right) + \frac{3 \pi^2 E_C}{2^{6} \kappa} \implies \kappa_{\text{optimal}} = \left[ \frac{2^{8} }{3\sqrt{3}\pi} \left( \frac{\Delta}{E_C} \right) \ln \left( 1+ \frac{E_C}{\Delta} \right) \right]^{-1/3}
\eeq
Hence, we finally find that the energy of optimal size skyrmion is given by 
\beq
E_{sk} = 4 \pi \rho_s + E_C \left( \frac{3^5 \pi^5 }{2^{13}\sqrt{3}} \right)^{1/3} \left[ \left( \frac{\Delta}{E_C} \right) \ln \left( 1+ \frac{E_C}{\Delta} \right) \right]^{1/3} \approx 4 \pi \rho_s + 1.75 \, E_C \left[ \left( \frac{\Delta}{E_C} \right) \ln \left( 1+ \frac{E_C}{\Delta} \right) \right]^{1/3}
\eeq
We immediately see that our analytical estimate of $E_{sk}$ in Eq.~(\ref{eq:EskApprox}) receives a logarithmic correction. For small Zeeman fields $B_\perp$ and intervalley coupling $\bro$, the energy of the skyrmion grows as $E_{sk}(\Delta) \approx [\Delta \ln (E_C/\Delta)]^{1/3}$. At larger fields (when the Zeeman energy becomes roughly of the order of the Coulomb energy) the size of the skyrmion will saturate, but an accurate estimate of the required magnetic field depends on lattice scale physics, and cannot be obtained from the low-energy field theory. 

Next, we turn to the effects of screening of the Coulomb interaction, which is relevant due to the metallic gates used on twisted bilayer graphene (see Eq.~(\ref{eq:ScrC})). Since the long-range (small $q$) nature of the Coulomb interaction is responsible for the $1/R$ scaling of the Coulomb energy with skyrmion size, we expect this scaling and thereby the optimal size and energy of the skyrmion to be significantly affected by screening. We assume that the gate-screened Coulomb interaction takes the following form discussed in Eq.~(\ref{eq:ScrC}). In the limit of small linear size of the skyrmion compared to the screening length $D$, i.e, $R \ll D$, screening effects are irrelevant and our previous result for the skyrmion energy holds (Eq.~(\ref{eq:EsAna})). However, the more relevant limit (where our continuum theory is likely to work better) is the large skyrmion size limit with $R \gg D$, as the screening length is typically of the order of a few moir\'e lattice spacings $a_M$. In this limit, the interaction term reduced to a short-range (contact-like) term. More precisely, the Fourier transformed charge density $\rho_\q$ is significant only for $q \lesssim 1/R$; in this regime $q D \ll q R $ and therefore $V_{\text{screened}}(\q)  \approx \frac{e^2 D }{2 \epsilon}$.  Using Eq.~(\ref{eq:Es}) and parametrizing $R = \kappa a_M$, we repeat the previous computations and find that our results for optimal size and energy are altered as follows for $D \approx a_M$ ($\alpha$ is an O(1) numerical constant). 
\beq
\kappa_{optimal} \propto  \left[ \left( \frac{\Delta}{E_C} \right) \ln \left( 1+ \frac{E_C}{\Delta} \right) \right]^{-1/2}, \text{ and } E_{sk} = 4 \pi \rho_s +  \alpha E_C \left[ \left( \frac{\Delta}{E_C} \right) \ln \left( 1+ \frac{E_C}{\Delta} \right) \right]^{1/2}
\label{eq:EsAnaScr}
\eeq
We note that the energy of the skyrmion grows as $E_{sk}(\Delta) \approx [\Delta \ln (E_C/\Delta)]^{1/2}$ as a function of the magnetic field in this case. Therefore, it is reasonable to expect that $E_{sk}(\Delta) \approx [\Delta \ln (E_C/\Delta)]^{\nu}$ for some $\nu \in (1/3,1/2)$ will accurately capture intermediate screening. Irrespective of the exact value of the exponent $\nu$, the estimate for the saturation lengthscale for the skyrmion remains identical, i.e, $\ell_{\tilde{B}} \approx \sqrt{a_0 a_M}$.

Finally, we discuss how the energetics of the skyrmion in a magnetic field are significantly different for a spin-valley locked state. In this case, the low energy Lagrangian density is given by:
\beq
\mathcal{L} = \sum_{\tau=\pm} \left[ n S \bigg( \bm{A}[\n_\tau] \cdot \partial_t \n_\tau(\r)  + g_s \mu_B \B \cdot \n_\tau(\r) \bigg) - \frac{\rho_s}{2} (\nabla \n_\tau (\r))^2 \right] - \frac{n S^2 \bro}{2} [(\n_+(\r) + \n_-(\r)]^2- \frac{1}{2} \int d\rp \; V(\r - \rp) \rho(\r) \rho(\rp) \nn
 \label{eq:LeffAFMApp}
 \eeq
In presence of a magnetic field $B_\perp$, the ground state is a canted antiferromagnet, with spins in each valley canting towards $B_\perp$. The optimal canting angle $\theta_0(B_\perp)$ can be obtained by minimizing the local energy for a spatially uniform ground state with $\n_+ = (\cos \theta, 0, \sin \theta), \; \n_- = (\cos \theta, 0, -\sin \theta)$. 
\beq E(\theta) &=& -\frac{g_s \mu_B}{2} \B \cdot(\n_+ + \n_-)  + \frac{\bro}{8}(\n_+ + \n_-)^2 = - g_s \mu_B B_\perp \sin \theta + \frac{\bro}{2} \sin^2 \theta ; \nn
\frac{\partial E}{\partial \theta}\bigg|_{\theta = \theta_0} &=& 0 \implies \sin \theta_0 = \begin{cases} \frac{g_s \mu_B B_\perp}{\bro}, \; B_\perp \leq \frac{\bro}{g_s \mu_B} \\
1, \text{   otherwise} \end{cases}
\eeq
We now find the effective magnetic field $\B_{\text{eff}}$ acting on the (ferromagnetic) spins of a single valley (say +), which will determine the magnon gap $\Delta$. We expect $\B_{\text{eff},+}$ to be parallel to the ferromagnetic order parameter $\n_{+}$ at equilibrium; we show that this is explicitly true below (taking $\hat{e}_\parallel$ and $\hat{e}_\perp$ to be the axes parallel and normal to $\n_+(\theta_0)$). 
\beq
\B_{\text{eff},+} = B_\perp \hz - \frac{\bro}{2 g_s \mu_B} \n_-(\theta_0) = \left( B_\perp \sin \theta_0 + \frac{\bro}{2 g_s \mu_B} \cos(2\theta_0) \right)  \hat{e}_\parallel + \left( B_\perp \cos \theta_0 - \frac{\bro}{2 g_s \mu_B} \sin(2\theta_0) \right) \hat{e}_\perp  = \frac{\bro}{2 g_s \mu_B}\hat{e}_\parallel \nn
\eeq
Therefore, the Zeeman gap for each valley is given by
\beq
\Delta = g_s \mu_B |\B_{\text{eff}}| = \begin{cases} \frac{\bro}{2}, \;  B_\perp < \frac{ \bro}{g_s \mu_B} \\  g_s \mu_B B_\perp - \frac{\bro}{2},\; B_\perp \geq \frac{ \bro}{g_s \mu_B} \end{cases}
\eeq
Therefore, we find that the unlike the ferromagnet, the Zeeman gap $\Delta$ initially remains fixed as the spins in each valley reorient in the ground state to give a canted antiferromagnet, and only starts to increase beyond a critical field of $B_c = \bro/(g_s \mu_B)$. This implies that the skyrmion size and the charge gap (due to charge $e$ skyrmions) also remains fixed till $B_c$. On further increasing $B_\perp$ beyond $B_c$, we get analogous behavior to the ferromagnet, as the skyrmion begins to shrink in size and increase in energy as $(B_\perp - B_c)^{\nu}$ with logarithmic corrections. 

\subsection{Skyrmion pairs}
\label{app:EnSkyPair}
In this section, we compute energy of skyrmion pairs, and discuss the situations where skyrmion pairing is favored at the lowest energy scales. First, let us consider the ferromagnet with $\langle s^z \rangle \neq 0$, and discuss pairing between skyrmionic charges in the same valley. This will be the case when the inter-valley coupling $J^\prime$ is much smaller than the intra-valley coupling $J$, as such a scenario will prefer the spins within the same valley to be aligned at the small cost of misalignment of spins in opposite valleys. For a charge $2e$ pair, we need the skymions to carry the same Pontryagin index but opposite phases. Therefore, consider the skyrmion pair ansatz given by:
\beq
W(z) = \frac{R}{z-L} - \frac{R}{z + L}
\eeq
The elastic energy for $W(z)$ with $2$ poles is $8 \pi \rho_s$, while the effective Zeeman energy is given by:
\beq
E^Z_{\text{pair}} = \frac{g_s \mu_B \tilde{B}}{\sqrt{3} a_M^2} \int_{0}^{\infty} d^2 r \; [1 - n_z(\r)] , \; \;  \text{ where, as before } \tilde{B} =  B_\perp + \frac{\bro}{g_s \mu_B}
\eeq
We now expect the logarithmic divergence to be cut off by $L$ instead of $\xi_s$, which we verify by an explicit calculation below. 
\beq
E^Z_{\text{pair}} &=& \frac{g_s \mu_B \tilde{B}}{\sqrt{3} a_M^2} \int_{0}^{\infty} dr \; r \int_0^{2\pi} d\theta \frac{2 (2 L R)^2}{r^4 - 2 L^2 r^2 \cos(2\theta) + D^4 + (2 L R)^2} \nn
& = & \frac{16 \pi g_s \mu_B \tilde{B} R^2 L^2}{\sqrt{3} a_M^2} \int_{0}^{\infty} dr \; r \frac{ 2 \pi}{\sqrt{(r^4 - L^4 +  4 D^2 R^2 )^2 + 16 L^6 R^2}} \nn
& \approx & \frac{8 \pi g_s \mu_B \tilde{B} R^2}{\sqrt{3} a_M^2} \ln \left( \frac{2L}{R} \right) \; \text{ for } \frac{R}{L} \ll 1 
\label{eq:EzPair}
\eeq

The Coulomb energy of interaction between the skyrmions (labeled $\pm$ according to their centers at $\pm L \, \hat{x}$) can be written down as:
\beq
E^\text{C}_{\text{pair}} &=&  e^2 \int \frac{d^2q}{(2 \pi)^2} V(\q) \rho_{+,\q} \rho_{-,-\q}  \text{ where } \rho_{\pm,\q} = \rho_\q e^{ \pm  i \q \cdot D \hx} \nn
& = & \frac{e^2}{4 \pi \epsilon} \int_0^\infty dq \; (qR)^2 \, [K_1(qR) ]^2 J_0(2 q L)
\label{eq:ECPair}
\eeq
The integral in Eq.~(\ref{eq:ECPair}) is cut off at $q \approx 1/L$ in the limit of small $R/L$ (skyrmion sizes are small compared to their separation), while for small separation $2L$ compared to the skyrmion size $R$ it is cutoff by $q \approx 1/R$.  Recall that $2L$ is the separation between the skyrmions, so in the limit of small $R/L$ we can write down the net energy of the skyrmion pair as follows (neglecting the self-Coulomb energy).
\beq
E_{\text{pair}} = E^{\text{elastic}}_{\text{pair}} + E^Z_{\text{pair}} + E^\text{C}_{\text{pair}} = 8 \pi \rho_s + \frac{8 \pi g_s \mu_B \tilde{B} R^2}{\sqrt{3} a_M^2} \ln \left( \frac{2L}{R} \right) + \frac{e^2}{4 \pi \epsilon (2L)} 
\label{eq:ESkPairApp}
\eeq
It is evident from Eq.~(\ref{eq:ESkPairApp}) that there is a minima in the energy at a finite separation $2L$, and therefore a bound state of two skyrmions will be formed. Minimizing $E_{\text{pair}}(L)$ in Eq.~(\ref{eq:ESkPair}) as a function of $L$, we find that $2L \approx \left(\frac{a_M}{R} \right)^2 \ell^2_{\tilde{B}}/a_0$ as the optimal separation between the skyrmions of size $R$. Since the inter-valley coupling $\bro$ is the smallest scale in the problem, the corresponding magnetic length $\ell_{\tilde{B}}$ will be very large and therefore our assumption of $L \gg R$ is self-consistent. We carefully note that the mean-separation $2L$ between the two skyrmions needs to be less than $\xi_s$, as at very large distances greater than $\xi_s$ only the repulsive Coulomb interaction, which disfavors pairing, operates \cite{NK_PRL98}. Recall that $\xi_s = \left( \frac{E_C}{\Delta}\right)^{1/2} a_M$, so such a regime always exists as long as the effective Zeeman energy is not too large. Further, as discussed in the main text (see also Ref.~\cite{NK_PRL98}), such a skyrmion pair carries spin, so the superconductor obtained by skyrmion-pairing also breaks spin-rotation (and time-reversal) symmetry. 

Skyrmionic charges pairing from opposite valleys need to have opposite Pontryagin indices so that they have the same physical charge (because of their opposite Chern numbers). There are two ways to do so: $\n \rightarrow -\n$ (which will cost a huge amount of energy in a large system as spins far away are antialigned) and $\n = (n_x,n_y,n_z) \rightarrow (-n_x,n_y,n_z)$ or $(n_x,-n_y,n_z)$, which will be relatively more favorable from energetic considerations. In either case, the skyrmion pair configuration does not lead to a gain in the effective Zeeman energy (unlike the previous scenario) as there is no quenching of the perpendicular components of the spin at distances larger than the skyrmion separation. Neither can it gain energy from alignment of spins in opposite valleys by having the two skyrmions sit on top of each other ($D \lesssim R$), as the requirement of opposite Pontryagin index forces the effective Zeeman energy to add up (it is approximately $2 \pi \bro (R/a_M)^2 \ln(\xi_s/R)$ in the continuum limit), and further, the Coulomb energy of placing two charges on top of each other also becomes large. Therefore, we conclude that there is no binding glue for skyrmions from opposite valleys in the ferromagnet. On the contrary, both Coulomb and Zeeman energy favors a charge-neutral skyrmion-pairing from opposite valleys, resulting in a time-reversal symmetry breaking intervalley coherent phase as discussed in the main text. 

Next, we turn to the spin-valley locked state. Once again, we start by discussing pairing between skyrmions in the same valley at zero external magnetic field ($B_\perp = 0$). Skyrmions with opposite phases still lead to an effective Zeeman energy (as $\tilde{B} \propto \bro \neq 0$) which is logarithmic in their separation for $D \gg R$. The energy of the skyrmion pair is given by
\beq
E_{\text{pair}} = E^{\text{elastic}}_{\text{pair}} + E^Z_{\text{pair}} + E^\text{C}_{\text{pair}} = 8 \pi \rho_s  + \frac{8 \pi \bro R^2}{\sqrt{3} a_M^2} \ln \left( \frac{2L}{R} \right) + \frac{e^2}{4 \pi \epsilon_0 (2L)} 
\label{eq:ESkPairAF}
\eeq
which is identical to Eq.~(\ref{eq:ESkPair}) for the ferromagnet at zero external magnetic field ($B_\perp = 0$). To summarize, the physics of pairing is analogous to the corresponding ferromagnetic case, and the skyrmion pair will also carry a large spin.  

Finally, we discuss the pairing between skyrmions in opposite valleys for the spin-valley locked state. In this case, skyrmion from one valley and an anti-skyrmion from the opposite valley can prevent any loss of inter-valley exchange energy by simply sitting on top of each other and locally satisfying $\n_+(\r) = -\n_-(\r)$. Such a configuration has twice the charge of a single-valley skyrmion, so its Coulomb energy goes as $1/R$ where $R$ is its size, and can be almost negligible for a large enough skyrmion-sizes. In the limiting case of $R \rightarrow \infty$, the energy of this skyrmion pair is simply $8 \pi \rho_s$. Such a skyrmion-antisykrmion pair thus avoids both the effective Zeeman energy cost by keeping spins from opposite valleys locally anti-aligned, and Coulomb energy cost by distributing the charge over a large lengthscale; it is the minimum energy skyrmion pair.

\end{document}